\documentclass[iop]{emulateapj}
\usepackage{amsmath}
\usepackage{longtable}
\usepackage{threeparttablex} 
\usepackage{hyperref}
\usepackage{color}

\shorttitle{BH Spin flips}
\shortauthors{Li et al.}
\newcommand{\msun}{M$_{\odot}$}

\begin{document}

\title{Secular Spin-orbit Resonances of Black Hole Binaries in AGN Disks}

\author{Gongjie Li}
 \email{gongjie.li@physics.gatech.edu}
 \altaffiliation{Center for Relativistic Astrophysics,
School of Physics,
Georgia Institute of Technology,
Atlanta, GA 30332, USA}
\author{Hareesh Gautham Bhaskar}%
\affiliation{%
 Center for Relativistic Astrophysics,
School of Physics,
Georgia Institute of Technology,
Atlanta, GA 30332, USA
}%
\author{Bence Kocsis}
\affiliation{
 Rudolf Peierls Centre for Theoretical Physics, University of Oxford, Parks Road, Oxford, OX1 3PU, United Kingdom
}%
\author{Douglas N. C. Lin}
\affiliation{%
Department of Astronomy and Astrophysics, University of California, Santa Cruz, CA 95064, USA
}%

\begin{abstract}

The spin-orbit misalignment of stellar-mass black hole (sBH) binaries provide important constraints on the formation channels of merging sBHs. Here, we study the role of secular spin-orbit resonance in the evolution of a sBH binary component around a supermassive BH (SMBH) in an AGN disk. We consider the sBH's spin-precession due to the $J_2$ moment introduced by a circum-sBH disk within the warping/breaking radius of the disk. We find that the sBH's spin-orbit misalignment (obliquity) can be excited via spin-orbit resonance between the sBH binary's orbital nodal precession and the sBH spin-precession driven by a massive circum-sBH disk. Using an $\alpha$-disk model with Bondi-Hoyle-Lyttleton accretion, the resonances typically occur for sBH binaries with semi-major axis of $1$AU, and at a distance of $\sim 1000$AU around a $10^7$\msun SMBH. The spin-orbit resonances can lead to high sBH obliquities, and a broad distribution of sBH binary spin-spin misalignments. However, we note that the Bondi-Hoyle-Lyttleton accretion is much higher than that of Eddington accretion, which typically results in spin precession being too low to trigger spin-orbit resonances. Thus, the secular spin-orbit resonances can be quite rare for sBHs in AGN disks. 
\end{abstract}


\section{Introduction} \label{sec:intro}
The detection of gravitational waves from stellar mass black hole (sBH) mergers provides an unprecedented opportunity to probe the properties of sBH binaries. For instance, the measurement of the spin-orbit misalignment and the spin precession reveal key information to better characterize the sBH binaries \citep{Cutler94, Chatziioannou15}, and to determine the origin of the sBHs \citep{Rodriguez16, Farr17, Gerosa17}. Hundreds of mergers of stellar mass compact binaries will be detected within the next decade, with the improved sensitivity of LIGO-Virgo-KAGRA and the expected commissioning of LIGO India. This prospect will enable detailed explorations of the origin and evolution of these compact binaries based on the statistical properties of the sBH binaries. 

The dynamics of sBH binary spin-orbit coupling is rich. For isolated binaries, spinning sBH can ``flip-flop'' when the binary components are close to each other during the merger process within a separation of $\sim 10^3 M$ \citep{Lousto15, Gerosa19}, where $M$ is the total mass in units $G=c=1$. Such occurrence leads to large sBH's spin-orbit misalignment up to $180^\circ$ between the sBH's spin and the orbital normal direction. Numerical-relativity simulations of sBH binaries revealed that spin-orbit misalignment can lead to large sBH recoils, with important astrophysical implications \citep{Campanelli07, Gonzalez07, Brugmann08, Kesden10, Lousto11, Gerosa18}. 

For sBH binaries around a supermassive black hole (SMBH), large sBH spin-orbit misalignment can be produced when the binaries are highly misaligned with respect to their orbits around the SMBH prior to the merger process. Specifically, the Von Zeipel-Lidov-Kozai mechanism due to the Newtonian perturbation of the SMBH can lead to inclination oscillation and eccentricity excitation of the sBH binary's orbit around each other and induce sBH merger \citep[see review by][]{Naoz16}. Meanwhile, the sBH spin undergoes chaotic evolution \citep{Liu17, Fragione20}, leading to a wide range ($0-180^\circ$) of final spin-orbit misalignment even from an initially aligned configuration. However, when the binary orbital inclination relative to its orbit around the SMBH is low ($\lesssim 40^\circ$), the binary eccentricity is not excited and only modest ($\lesssim 20^\circ$) spin-orbit misalignment can be produced.

Here, we show that in AGN disks, spin-orbit resonance can increase the sBHs' obliquities even for sBH binaries in near coplanar orbits around the SMBH. AGN disks are dense with stars and compact objects \citep{Bartos17,Tagawa20}, and such environments are conducive to forming and merging of compact-object binaries. 
How do the spins of the sBHs evolve? \citet{McKernan20} adopted Monte Carlo simulations and found the distribution of the effective spin parameter $\chi_{\rm eff}$ is  centred around $\tilde{\chi }_{\rm eff} \approx 0.0$ with a width of the $\chi_{\rm eff}$ distribution for low natal spins. In addition, \citet{Tagawa_spin20} found that the strong binary-single encounters can randomize the orbital inclination and lead to non-zero but equal obliquity for the binary components due to Bardeen-Petterson effect and binary-single hard encounters. However, effects of secular spin-orbit resonances of the sBHs (gentle but correlated perturbations which accumulate over many orbital periods) in the AGN disk have not been studied. 

In this article, we propose a new mechanism to change the spin orientation of the sBHs via secular spin-orbit resonances. Secular dynamics on the eccentricity increase and the enhancement of merger rate of sBH binaries in AGN disks have been recently investigated in \citet{Bhaskar22, Bhaskar23N}. Here, we consider the spin dynamics. Specifically, inner part of the circum-sBH disk that strongly couples to the spin of the sBH and is driven by Lense-Thirring precession, and the Newtonian torque acting on the disk by the companion sBH leads to retrograde spin-precession. We show that for massive circum-sBH disk with Bondi-Hoyle-Lyttleton accretion, the retrograde spin precession can dominate over the prograde de Sitter spin precession, and resonate with nodal precession of the binary orbit driven by the perturbation of the SMBH. The spin-orbit resonance can lead to highly misaligned sBH spins orbiting closely ($\sim 10^3$ AU) around SMBHs ($10^7M_{\odot}$). The tilting of sBH due to spin-orbit resonance is analogous to that of the tilting of Saturn including the effects of satellites and circumplanetary disk \citep{Ward04, Rogoszinski20}. This can broaden the distribution of spin-spin misalignment and $\chi_{\rm eff}$. 

\section{Spin Variations}
\label{sec:spin-orbitres}

\subsection{System setup}
\begin{figure}
\center
    \includegraphics[width=0.45\textwidth]{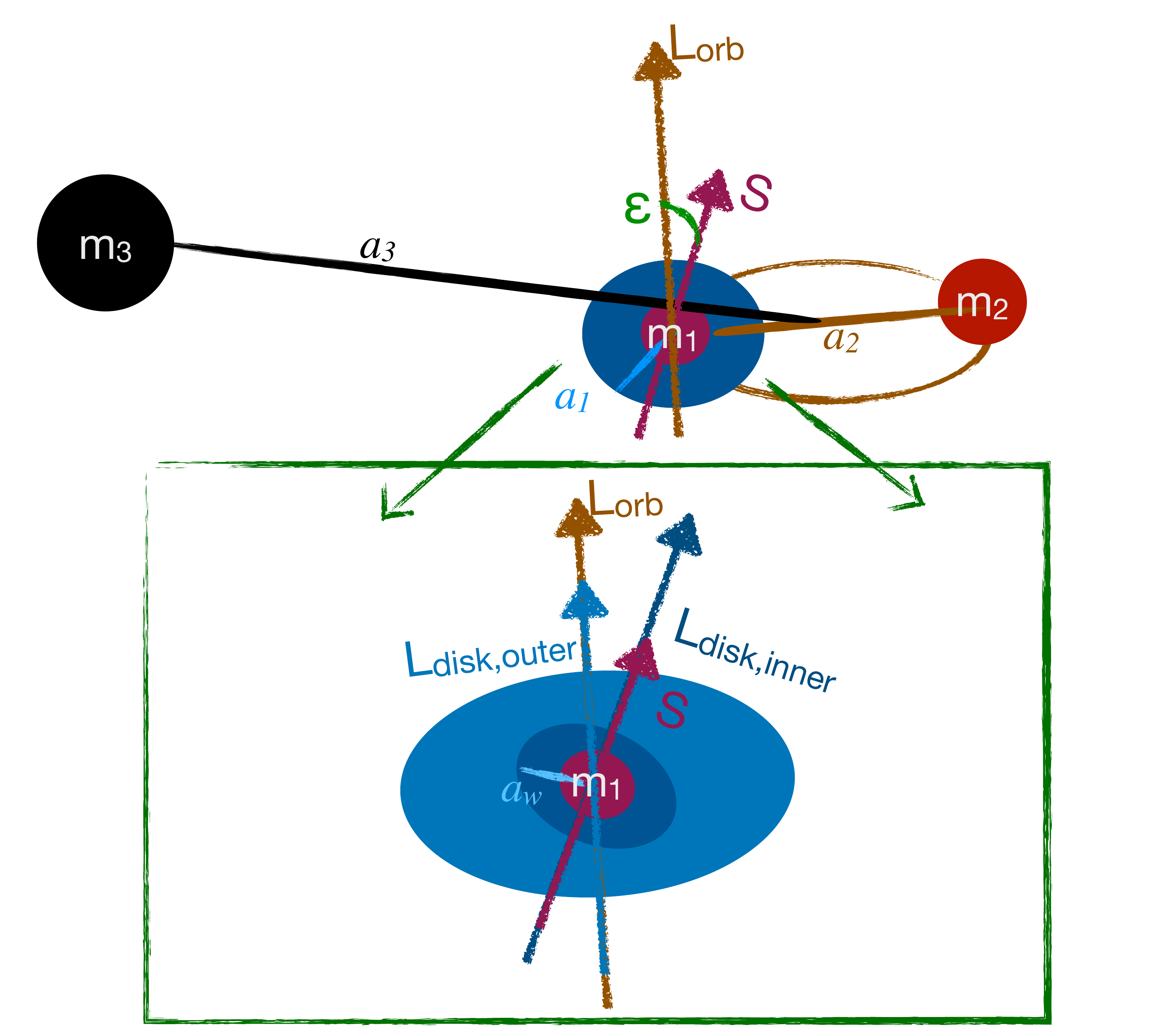}
    \caption{Configuration of the system. The sBH binary is composed of $m_1$ and $m_2$, and $m_3$ represents the SMBH. The blue region represents the gaseous disk around the central sBH $m_1$. $a_1$ is the semi-major axis of the disk particles around $m_1$, $a_2$ is the semi-major axis of the sBH binary and $a_3$ is the semi-major axis of the sBH binary around the SMBH. The green $\epsilon$ angle represents the sBH's spin-orbit misalignment (obliquity). The green box shows a zoom in view of the warped disk, where inside the warping radius $a_w$, the disk is strongly coupled to the sBH. The warping radius is located close to the Laplace radius (eqn \ref{eqn:L}). \label{fig:config}}
\end{figure}

We examine the spin variations of sBH binaries embedded in a SMBH AGN disk. The configuration is shown in Figure \ref{fig:config}. The binary is composed of sBH ($m_1$) and sBH ($m_2$), and $m_3$ represents the SMBH. The blue region around $m_1$ represents the circum-sBH disk. The total spin angular momentum of the sBH and the inner circum-sBH disk is denoted by ${\bf S}$. The orbital angular momentum of the sBH binary is marked by ${\bf L}_{orb}$. The semi-major axis of the disk particles is denoted by $a_1$. $a_2$ is the semi-major axes of the sBH binary, and $a_3$ denotes the semi-major axis of the orbit of the sBH binary around the SMBH ($m_3$). The green box in Figure \ref{fig:config} shows a zoom-in view of the circum-sBH disk. The spin-orbit misalignment is marked using $\epsilon$ in the figure. We note that ${\bf L}_{orb}$ and ${\bf L}_{disk,innner}$ are not aligned when the sBH spin is misaligned with the orbit (${\bf L}_{orb}$).



\subsection{Warping of circum-sBH disks}
\label{sec:warp}
When the spin of the sBH is misaligned with the binary orbit, the disk becomes warped (as illustrated in the zoom-in view in Figure \ref{fig:config}). Specifically, the inner region of the circum-sBH disk aligns with the sBH spin due to Bardeen-Petterson effect \citep{Bardeen75, papaloizou1983}, and while the outer parts of the disk align with the sBH binary orbital axis \citep[e.g.,][]{Martin09, Tremaine14}. This warping of the disk is analogous to the warping of the orbital plane of satellites, which is due to the precession of a satellite around its host planet and perturbed by the star \citep{Goldreich66}. We use $a_w$ to denote the warping radius, which marks the limit where the precession direction of the disk changes significantly. Tears of the disks can occur when the warp is significant (e.g., $\gtrsim 40^\circ$ depending on the disk properties), and this hinders the subsequent realignment of the sBH spin with the outer disk  \citep[e.g.,][]{Nixon12, Gerosa20, Nealon22}.

The viscosity of the disk determines whether the breaking occurs (also illustrated in \citet{pringle1992, ogilvie1999}), and the breaking radius is close to the Laplace radius. Specifically, the Laplace radius is located where the direction of the disk particle's nodal precession changes, and it can be estimated by setting the Lense-Thirring precession rate to be equal to the nodal precession rate. The nodal precession of the disk particles due to the spinning sBH can be expressed as follows \citep{Barker74}:
 \begin{align}
     \Big(\frac{d\Omega}{dt}\Big)_{LT} = \frac{2 G^2 m_1^2 \chi}{c^3a_1^3(1-e_1^2)^{3/2}},
 \end{align} 
where $e_1$ stands for the eccentricity of the orbit of the disk particle around sBH $m_1$, and $\chi$ stands for the spin parameter of the sBH $m_1$ (see eqn 5). Meanwhile, the companion sBH drives nodal precession around the orbital normal of the black hole binary. To the quadrupole order in the semi-major axes ratio ($a_1/a_2$), the precession rate is simple in the near coplanar limit \citep{Naoz16, Liu17}:
 \begin{align}
     \Big(\frac{d\Omega}{dt}\Big)_{vZLK} &
      = -\frac{3 n m_2}{4(m_1)} \Big(\frac{a_1}{a_2\sqrt{1-e_2^2}}\Big)^3 , \label{eqn:L}
 \end{align}
where $n = \sqrt{G(m_1+m_2)/a_2^3}$ is the mean motion of the binary, and $e_2$ stands for the eccentricity of the sBH binary.

Then, we obtain the Laplace radius ($a_L$) by equating $\Big(\frac{d\Omega}{dt}\Big)_{LT}$ and $\Big(\frac{d\Omega}{dt}\Big)_{vZLK}$, assuming the eccentricities of the disk particles and that of the sBH binary to be zero, and omitting factors of order unity:
\begin{align}
    a_{L} = a_2^{2/3} R_g^{1/3}\Big(\frac{\chi m_1}{m_2}\Big)^{2/9} , \label{eqn:L}
\end{align}
where $R_g = \frac{ G m_1} {c^{2}}$ is the gravitational radius of $m_1$. This expression is within an order of unity of the warping radius obtained in \citet{Martin09, Tremaine14}. We denote it as the Laplace radius instead of the warping radius here, to distinguish it from the actual location of the disk where warps/breaks occur.

\begin{figure}
\center
    \includegraphics[width=0.45\textwidth]{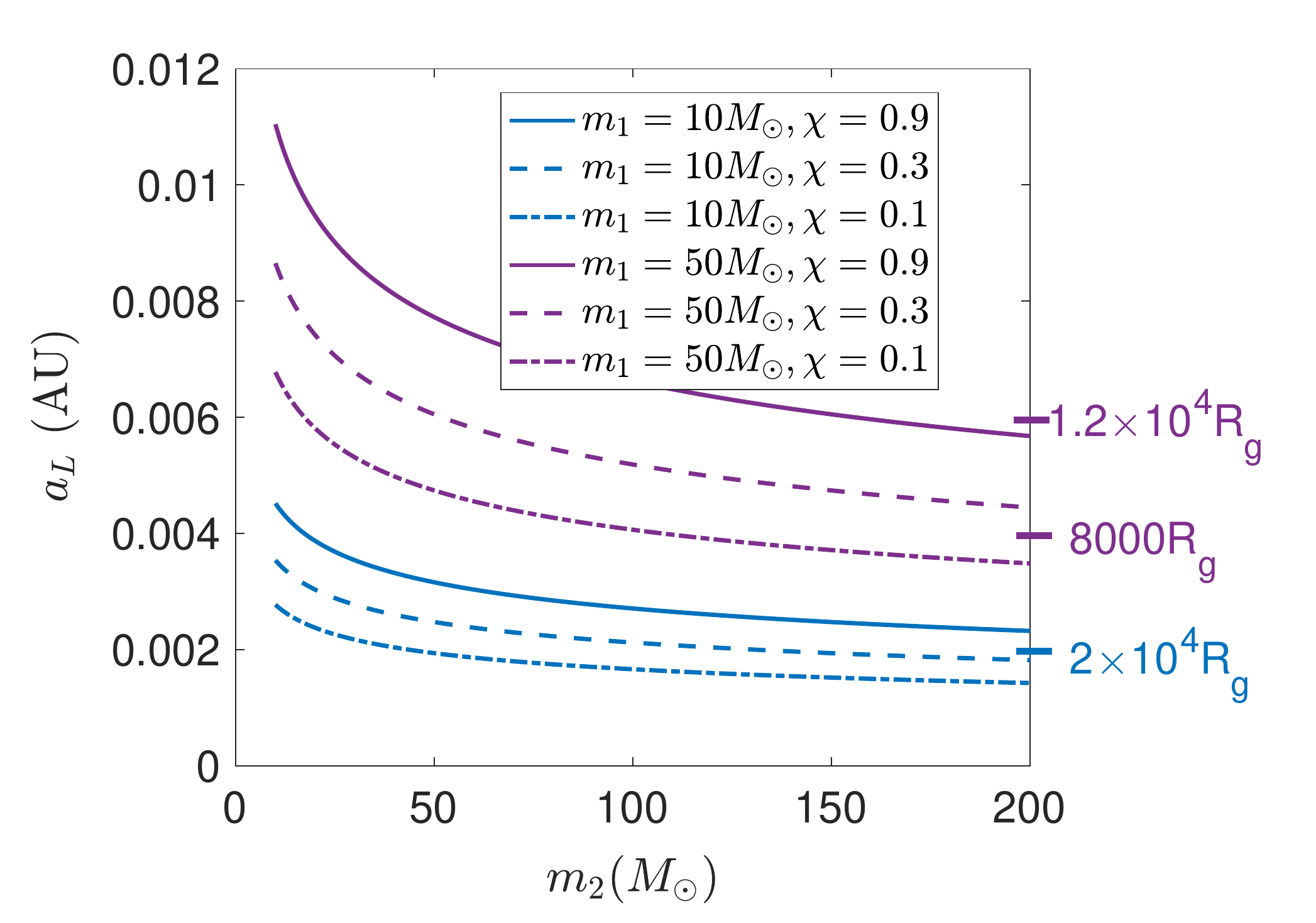}
    \caption{Laplace radius of disk particles around a $10$ \msun~and $50$ \msun~central sBH with sBH binary semi-major axis of $a_2=1$AU. Different types of lines represent different sBH spin coefficients. The y-axis label on the right shows the Laplace radius in terms of the gravitational radius of the central sBH. The blue lines and labels correspond to that around a $m_1=10$\msun~sBH, and purple corresponds to a $m_1=50$ \msun~sBH. \label{fig:laplace}}
\end{figure}

Figure \ref{fig:laplace} shows the Laplace radius of disk particles around a $m_1=10$ \msun~and $50$ \msun~central sBH. The semi-major axis of the sBH binary components is $a_2=1$AU, and we vary the mass $m_2$ of the sBH companion on the bottom-axis. As expected, the Laplace radius decreases with sBH companion mass, and the Laplace radius increases with the central sBH spin. At a sBH binary semi-major axis of $a_2=1$ AU, $a_{L}$ can greatly exceed $R_g$, as shown in the right vertical axis of the figure. The blue lines and labels represent that for the $10$ \msun~central sBH, and the purple represents the $50$ \msun~central sBH. The Laplace radius is smaller for smaller sBH separation ($a_2$).

Recently, it has been shown in 3-dimensional hydrodynamical simulations that significant warping as well as breaking can indeed occur for sBH spins misaligned with the binary orbit \citep{Nealon22}. The exact location of the breaking radius depends on the obliquity of the sBH spin, where retrograde spins allow for larger breaking radius. Overall, as shown in the 3-dimensional hydrodynamical simulations by \citet{Nealon22}, the breaking radius, where the discs separate into discontinuous structures, is about five times smaller than the Laplace radius depending on the inclination of the sBH spin axis. Accordingly, we set the warping/breaking radius ($a_w$) in terms of the Laplace radius ($a_L$) for rough estimates in the following discussions.


\subsection{Spin precession with circum-sBH disks}
\label{sec:spin_prec}

In this section, we consider the spin precession of the central sBH, taking into account both the de Sitter precession of the sBH, due to the curved spacetime produced by its sBH companion, as well as the precession of the disk due to the Newtonian tidal force of the companion. 

\subsubsection{Disk induced $J_2$ precession}
The disk around the central sBH produces a $J_2$ quadrupole moment (see eqn \ref{eqn:J2}), and the torque due to the sBH companion on the $J_2$ moment can lead to spin precession. The torque be expressed in the secular limit as follows
\citep{Goldreich66}:
\begin{align}
    \Big(\frac{d\hat{\bf s}}{dt}\Big)_{disk} &= - \frac{3Gm_2 J_2 m_1 R^2}{a_2^3 L_{spin}}(\hat{\bf l} \cdot \hat{\bf s})(\hat{\bf l} \times \hat{\bf s}) ,
    \label{eqn:precdisk}
\end{align}
where $R$ can be defined as the gravitational radius of the sBH, $\hat{\bf s} = {\bf L_{spin}}/L_{spin}$ is the unit vector pointing in the direction of the spin, and $\hat{\bf l}$ is the unit vector along the orbit normal. $L_{spin}$ is the total angular momentum of the sBH and the close-in disk that strongly coupled to the sBH spin ($\bf{L_{spin}} = \bf{L_{spin, sBH}} + \bf{L_{spin, disk}}$). The angular momentum of the sBH can be expressed as follows: 
\begin{align}
    L_{spin, sBH} &= \frac{Gm_1^2\chi}{c} .
\end{align}
The angular momentum of the disk depends on the circum-sBH disk density profile. Using a simple $\alpha$ disk model, where the surface density of the disk can be expressed as follows \citep{shakura_sunyaev_1973, FrankKing02}:
\begin{align}
    \Sigma (r) &= 5.2{\rm g/cm^{-2}}\alpha^{-4/5}\dot{m}_{16}^{7/10}(m_1/M_{\odot})^{1/4}r_{10}^{-3/4}f^{14/5} 
    \label{eq:sigma}\\
    f &= \Big(1-\Big(\frac{r_{in}}{r}\Big)^{1/2}\Big)^{1/4} \\
    r_{10} &= r/(10^{10} {\rm cm})\\
    \dot{m}_{16} &= \dot{m}/(10^{16} {\rm g/s})
\end{align}
where $\alpha$ is the viscosity parameter, $\dot{m}$ is the accretion rate, $r$ is the distance of the disk particles from $m_1$. We assume $\alpha = 0.01$ for the circum-sBH disk, where the viscosity is likely dominated by turbulent viscosity. This is typical for disks around sBHs or neutron stars \citep[e.g.,][]{FrankKing02}.

In addition, we assume the accretion rate of gas in the circum-sBH disk is given by the Bondi–Hoyle–Lyttleton rate:
\begin{align}
    \dot{m} &=4 \pi r_{w} r_h \rho_{gas} \sqrt{c_s^2 + v_k^2} , \label{eq:mdot}\\
    v_k &= \sqrt{\frac{G (m_1+m_2)}{a_2}}\frac{m_2}{m_1+m_2} ,
\end{align}
where $r_{w}$ is width of the gas bound to the sBH ($m_1$), which is determined by the minimum of the Bondi-Hoyle-Lyttleton radius $r_{BHL} = G(m_1)/(c_s^2 + v_k^2)$ and Hill radius of $m_1$ ($r_{Hill} = a_2 (m_1 / 3 m_3)^{1/3}$, and $r_{h}$ is height of gas bound to $m_1$, which is determined by the minimum of $r_{w}$ and the scale height of the AGN disk ($H$). Sound speed ($c_s$) and disk gas density ($\rho_{gas}$) are determined using the properties of the AGN disk.

\begin{align}
    \Sigma_{AGN} = &\frac{c_s \kappa}{\pi G Q}  = {h m_3 \over {\sqrt 2} \pi Q a_3^2} \label{eqn:Sigma_AGN}\\
    \rho_{gas} = & \frac{\Sigma_{AGN}}{2 H}  = {m_3 \over {\sqrt 8} \pi Q a_3^3},
\end{align}
where $c_s = {H\Omega}/{\sqrt{2}}$, $h=H/r$, $\Omega^2 = {G m_3}/{a_3^3}$ and $\kappa = \Omega$ are the mid-plane sound speed, aspect ratio, angular and epicycle frequencies respectively. We set $h = 0.01$ according to reverberation mapping \citep{Starkey23}. Specifically, \citet{Starkey23} measured a $h$ on the order of $0.03$, and we reduced $h$ further, because reverberation mapping corresponds to the measurement of the photosphere at grazing angle, and thus the measured height is larger than the scale height \citep{Garaud07}.

We adopt the Bondi-Hoyle-Lyttleton (BHL) accretion here, assuming the super-Eddington accretion occurs episodically, during which the spin-orbit resonances can take place. The episodic accretion takes place during the viscous timescale $t_{vis} \sim r^2/\nu = 1/(\alpha\Omega_{m1}h^2)\sim 1$Myr, where $\Omega_{m1}$ is the Keplerian angular velocity around the sBH ($m_1$), and this provides an upper bound on the timescale of the flip of the sBH due to spin-orbit resonances. Note that at $a_3 \sim 10^3$AU from an $m_3 =10^7 M_\odot$ SMBH, the BHL accretion rate is much higher ($\gtrsim 10^4$ times higher) than the Eddington accretion rate. 
With high accretion rate, the strong radiation pressure can suppress further accretion onto the sBH and may reach an accretion-wind equilibrium, which may lead to a more massive disk (Ali-Dib et al., in prep). 
Although it gives $Q \gtrsim 1$ in the inner region of circum-sBH disk, within $a_L$, which still allows the disk to be stable against gravitational collapse, the total disk mass within $r_{BHL}$ and $r_{Hill}$ can be comparable to or above the mass of the sBH. Thus, we only use it as an upper bound to investigate the effects secular spin-orbit resonances in massive disks. 

To compare with an Eddington-limited accretion, we also calculate the surface density of the mini circum-sBH disk using Eddington-limited accretion: 
\begin{align}
    \dot{m}_E &=1.26\times10^{31} {\rm W} \frac{m_1/M_{\odot}}{c^2 \eta} , \label{eq:mdotE}
\end{align}
where we set the radiative efficiency to be $\eta = 0.01$ for an upper bound on the accretion rate. 

\begin{figure}[h]
\center
    \includegraphics[width=0.45\textwidth]{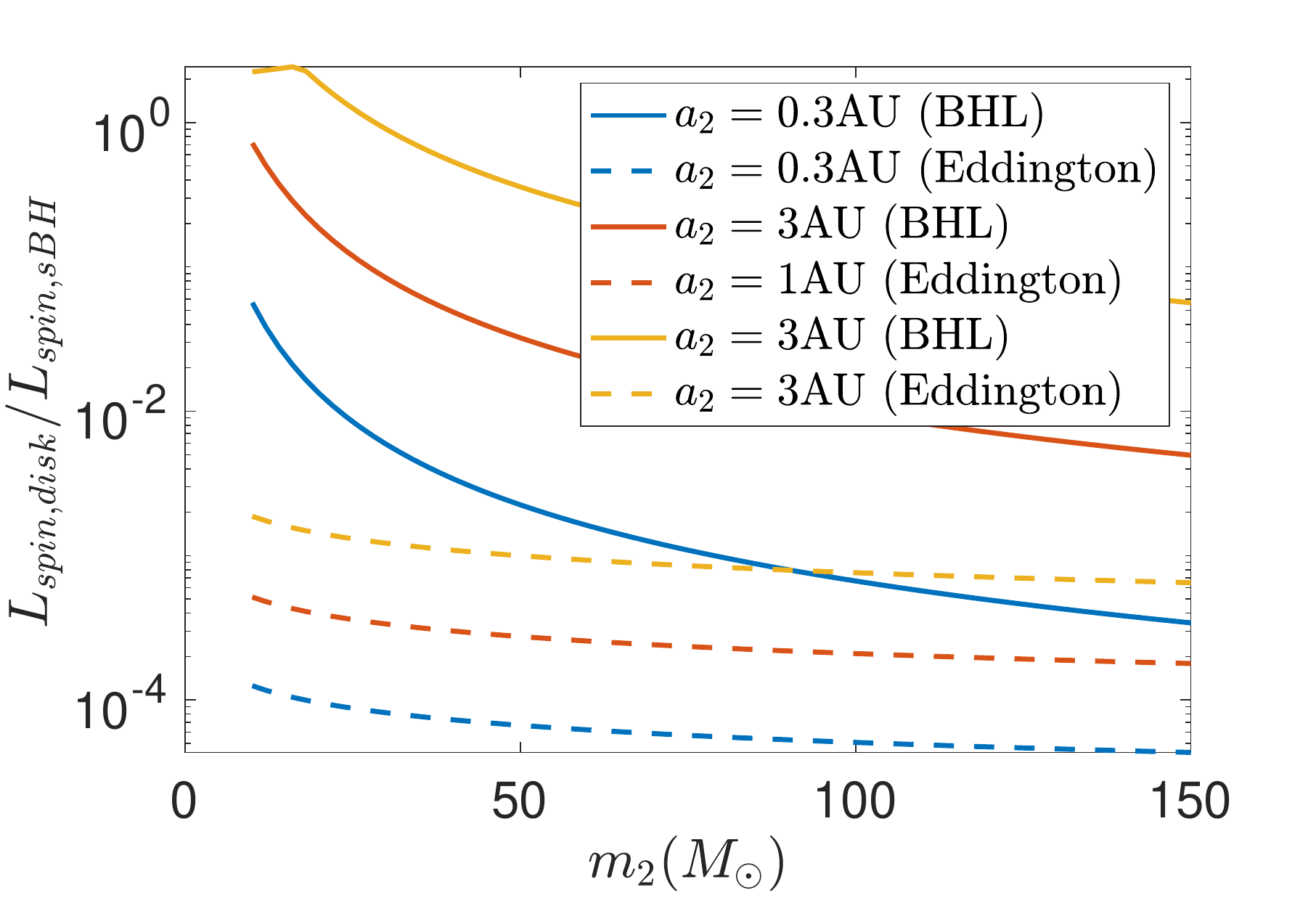}\\
    \includegraphics[width=0.45\textwidth]{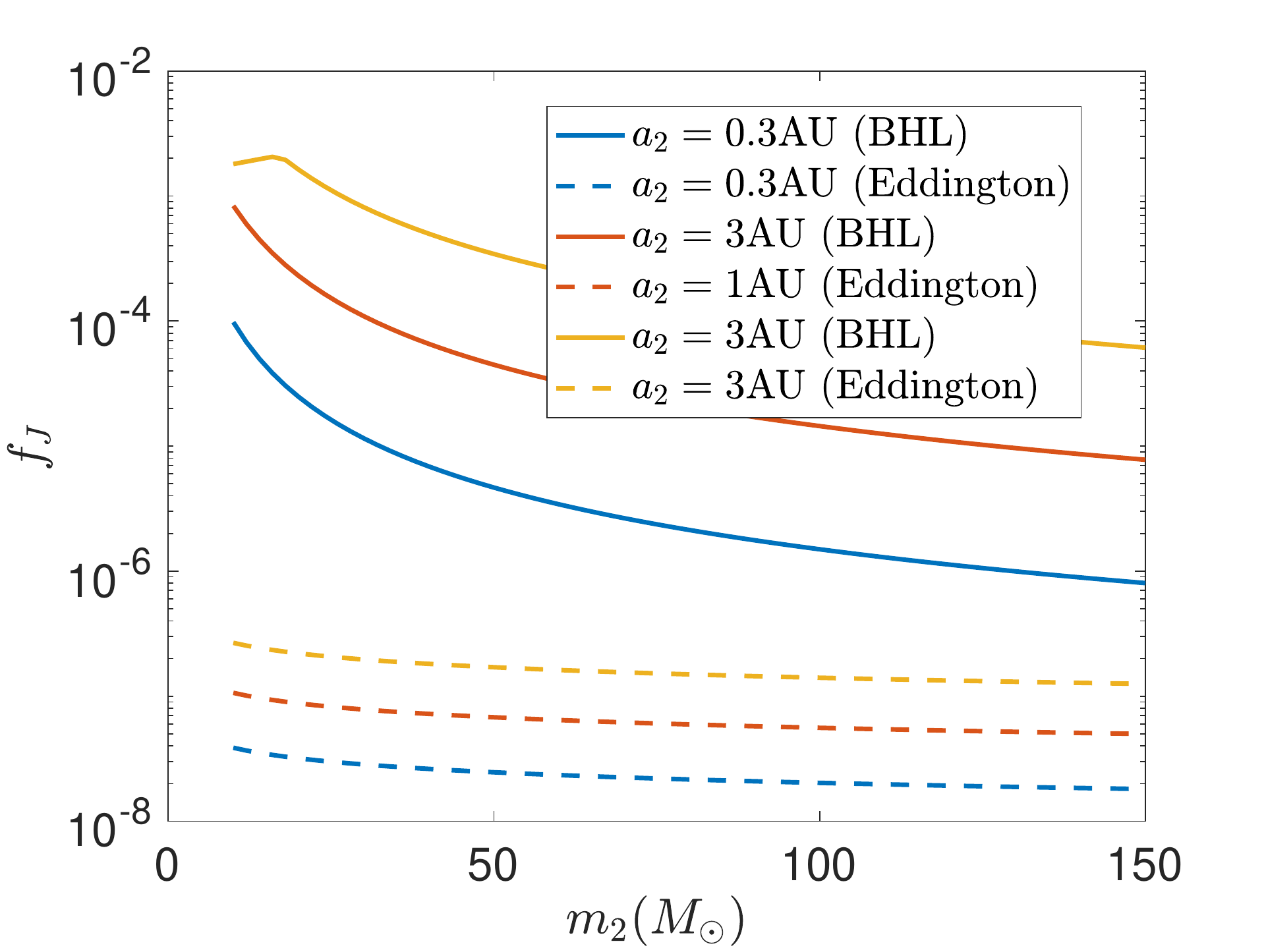}
    \caption{\textbf{Upper panel:} The disk angular momentum to sBH spin angular momentum ratio. \textbf{Lower panel:} The \textbf{$f_J$} parameter of the inner disk strongly coupled to the sBH spin. $f_J$ corresponds to the quadrupole moment of the inner disk ($f_J = \int_{a_{in}}^{a_{w}} \pi \Sigma(r) r^3 dr/(m_1 a_L^2)$. The Eddington-limited disk leads to much lower quadrupole moment than the BHL model.} \label{fig:LorbLBH}
\end{figure}

Following the disk surface density (Eqs. \ref{eq:sigma}, \ref{eq:mdot} and \ref{eq:mdotE}), the angular momentum of the disk can be calculated as follows:
\begin{align}
    L_{spin,disk} &= \int_{a_{in}}^{a_{L}} \Sigma(r)  2\pi r^3 \omega(r)dr ,
\end{align}
where $\omega=(G m_1/r^3)^{1/2}$. 

To illustrate the angular momentum caused by the circum-sBH disk and the sBH spin, we show the ratio of $L_{spin, disk}$ to $L_{spin, sBH}$ in the upper panel of Figure \ref{fig:LorbLBH}. We set $m_1 = 30$M$_\odot$, and $\chi = 0.7$ for the black holes, and we set the inner edge of the disk ($r_{in}$) to be $3R_g$. The sBH binary orbit around a SMBH of $10^7$\msun at $a_3 = 1000$AU. The blue line represents the case with $a_2 = 0.3$AU, the red one corresponds to $a_2 = 1$AU and the yellow line corresponds to $a_2 = 3$AU. The x-axis represents the mass of the sBH companion. The solid lines correspond to the BHL accretion disk model, and the dashed lines correspond to the Eddington-limited accretion.

We find that larger sBH binary separation corresponds to larger disks and thus the disk angular momentum increases with sBH separation. The larger the companion, the smaller the disk and thus the lower the angular momentum ratio. In addition, it shows that the angular momentum caused by the disk that strongly couples to the black hole spin is typically lower than that of the sBH spin when $a_1 \lesssim 3$AU. This is opposite to what is usually assumed for Bardeen-Petterson effects, since here we only consider the inner region of the disk strongly coupled to the Lense-Thirring precession of the black hole. The Eddington limited disk produces lower disk angular momentum.

Then, we calculate the quadrupole moment ($J_2$) introduced by the disk. It can be expressed as follows, similar to planets in a circumplanetary disk \citep[e.g.,][]{Rogoszinski20}: 
\begin{align}
    J_2 &= \frac{\int_{a_{in}}^{a_{L}} \pi \Sigma(r) r^3 dr}{m_1 R^2} , \label{eqn:J2}
\end{align}
where $R$ can be defined as the gravitational radius of the sBH. We express the integral ($\int_{a_{in}}^{a_{L}} \pi \Sigma(r) r^3 dr$) as $f_J m_1 a_L^2$, and $R$ can be canceled out in the expression of the spin precession. Specifically, 
\begin{align}
    f_J &= J_2 \frac{R^2}{a_L^2} ,
\end{align}
Thus, $d\hat{\bf s}/dt$ can be simplified as follows:
\begin{align}
    \Big(\frac{d\hat{\bf s}}{dt}\Big)_{disk} &= - \frac{3 f_J c m_2 a_L^2}{a_2^3 m_1 \chi}(\hat{\bf l} \cdot \hat{\bf s})(\hat{\bf l} \times \hat{\bf s}) .
    \label{eqn:diskspin}
\end{align}
Please note that a very crude estimate based on the definition of Toomre Q value ($Q=c_s \kappa/\pi G \Sigma \sim h m_1/(\sqrt{2}\pi\Sigma a_3^2)$, analogous to equation \ref{eqn:Sigma_AGN}) gives $f_J \sim h/Q$ at the Laplace radius.

To illustrate the magnitude of $f_J$, we show in the lower panel of Figure \ref{fig:LorbLBH} $f_J$ as a function of the sBH companion mass. Similar to the upper panel, the different colors represent different $a_2$. $f_J$ is larger for wider sBH separations, since the disks are larger. x-axis represents the mass of the sBH companion, and larger $m_2$ corresponds to lower $f_J$ since the disks are smaller. The Eddington-disk model (dahsed lines) leads to much lower $f_J$ comparing to the BHL disk model (solid lines), due to small amount of disk mass enclosed within the Laplace radius.

\subsubsection{De Sitter precession}
In addition to sBH spin precession due to the disk quadrupole moment, the sBH companion's spacetime curvature can drive sBH spin precession (de Sitter precession). The precession rate can be expressed as follows assuming the eccentricity is zero\citep{Barker75}:
\begin{align}
    \Big(\frac{d\hat{\bf s}}{dt}\Big)_{ds} &= \frac{3 G (m_2 + \mu/3)n }{2 c^2 a_2}(\hat{\bf l} \times \hat{\bf s}) ,
    \label{eqn:dsspin}
\end{align}
where $\mu = m_1 m_2/(m_1 + m_2)$ is the reduced mass of the sBH binary. Note that the de Sitter precession is opposite in direction to the Newtonian disk precession and the orbital precession. Thus, de Sitter precession does not lead to spin-orbit resonance, and spin-orbit resonances occur when the disk precession dominates over de Sitter precession. 

\begin{figure}[h]
\center
    \includegraphics[width=0.45\textwidth]{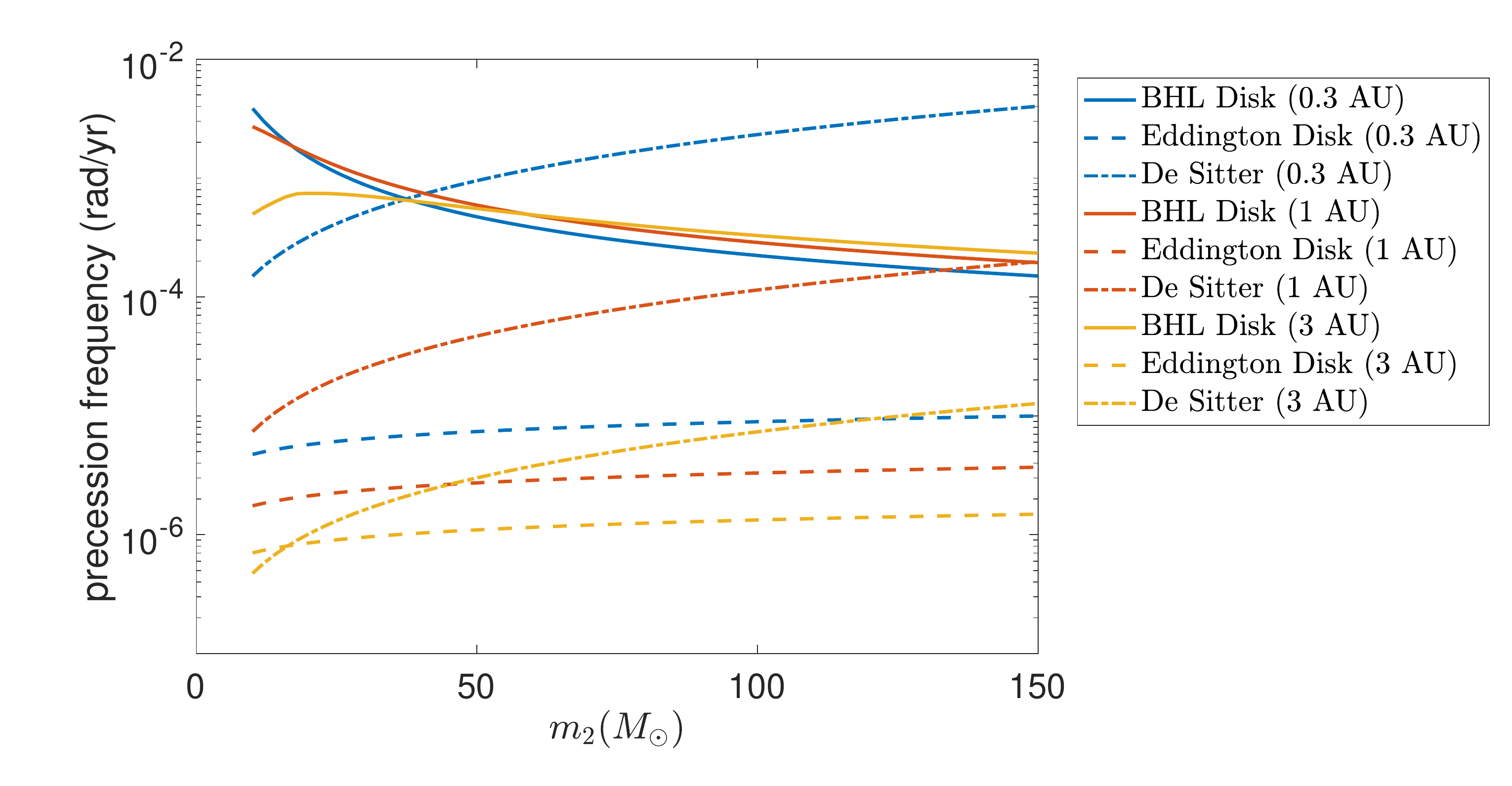}\\
    \caption{Spin precession rate due to de Sitter v.s. that due to Newtonian torque acting on the disk $J_2$ moment. Disk-induced $J_2$ precession dominates when the primary sBH is larger with farther sBH binary separation ($a_2 \gtrsim 1$AU), adopting the BHL disk model. Disk-induced precession following the Eddington-limited disk model is too low, comparing to the de Sitter precession.}  \label{fig:prec}
\end{figure}

To compare the two precession rates, we show in Figure \ref{fig:prec} their magnitudes for $m_1=30$\msun with different sBH companion masses. We set $e_2$ to be zero, and we assume that the spin is nearly aligned with the orbit ($\hat{\bf l} \cdot \hat{\bf s}\sim 1$). For illustration, we set $\chi = 0.7$. The precession rate due to the disk is larger for sBHs with lower spin coefficients. The disk precession timescale decays as the companion mass increases, since the disk size is smaller with larger companion. On the other hand, the de Sitter precession timescale increases with companion mass due to stronger space-time curvature. Thus, the disk precession dominates over the de Sitter precession for lower mass sBH companions.

In addition, we included three different black hole separations, similar to Figure \ref{fig:LorbLBH}. The precession due to the circum-sBH disk dominates over de Sitter precession only for wide sBH separations ($a_2 \gtrsim 1$AU), since the de Sitter precession drops faster with $a_2$. The solid lines correspond to the super-Eddington Bondi-Hoyle-Lyttleton accretion disk. With a lower surface density disk (e.g., in the Eddington limit as shown in dashed lines), the disk $J_2$ ponential is not large enough to allow disk precession to dominate over de Sitter precession. Thus, it is more challenging to induce spin-orbit resonances with lower surface density disks.

\section{Orbital Precession}
\label{sec:orb}
Spin-orbit resonance occurs when the spin precession frequency is close to that of the orbit. How does the sBH binary orbit precess? First, the sBH binary orbit precesses due to the $J_2$ potential of the disk and the Lense-Thirring precession \citep{Barker75}. Assuming eccentricities to be zero for simplicity, we get: 
\begin{align}
    \Big(\frac{d\hat{\bf l}}{dt}\Big)_{LT} =& \Big(\frac{G S_1 (4+3m_2/m_1)}{2c^2 a_2^3}(\hat{\bf s}_1 - 3(\hat{\bf l}\cdot\hat{\bf s}_1)\hat{\bf l}) \nonumber \\
    & + \frac{G S_2 (4+3m_1/m_2)}{2c^2 a_2^3}(\hat{\bf s}_2 - 3(\hat{\bf l}\cdot\hat{\bf s}_2)\hat{\bf l})\Big) \times \hat{\bf l}, \\
    \Big(\frac{d\hat{\bf l}}{dt}\Big)_{J2} = & -\frac{3}{2}n_2 f_{J1} \Big(\frac{a_{L,1}}{a_2}\Big)^2 (\hat{\bf s}_1 \cdot \hat{\bf l})(\hat{\bf s}_1 \times \hat{\bf l}) \nonumber \\
    &-\frac{3}{2}n_2 f_{J2} \Big(\frac{a_{L,2}}{a_2}\Big)^2 (\hat{\bf s}_2 \cdot \hat{\bf l})(\hat{\bf s}_2 \times \hat{\bf l}).
\end{align}
Here, we consider the spin of both of the sBH binary components, and subscript $1$ and $2$ denote that properties of $m_1$ and $m_2$ separately. Note that we neglected the spin-spin coupling, since the orbital angular momentum is much larger than that of the spin at a separation of $a_2 \sim 1$AU.

In addition, orbiting around the SMBH ($m_3$ in Figure \ref{fig:config}), the binary precesses due to the Newtonian perturbation of $m_3$. Assuming the inclination of the binary relative to the orbit around $m_3$ to be low, the nodal precession of the orbit is simple and can be expressed as follows \citep{Naoz16}:
\begin{align}
    \Big(\frac{d\hat{\bf l}}{dt}\Big)_{vZLK} = &
    -\frac{3 n_2 m_3}{4(m_1+m_2)} \Big(\frac{a_2}{a_3\sqrt{1-e_3^2}}\Big)^3 (\hat{\bf l}_3 \times \hat{\bf l}).
    \label{eqn:vZLK}
\end{align}
Note that we neglect the precession due to the circum-SMBH disk and the disk around the sBH binary. The disks produce retrograde sBH binary nodal precession, so the effects are qualitatively similar. Nevertheless, we assume that the precession due to the SMBH dominates as the SMBH mass is much more massive than the disk. In addition, we note that the orbital Lense-Thirring precession rate and the $J_2$ precession rate are much lower than the spin precession rate due to de Sitter precession and $J_2$ separately. Thus, we only consider the orbital precession due to the SMBH here. 

 \begin{figure}[h]
 \center
     \includegraphics[width=0.45\textwidth]{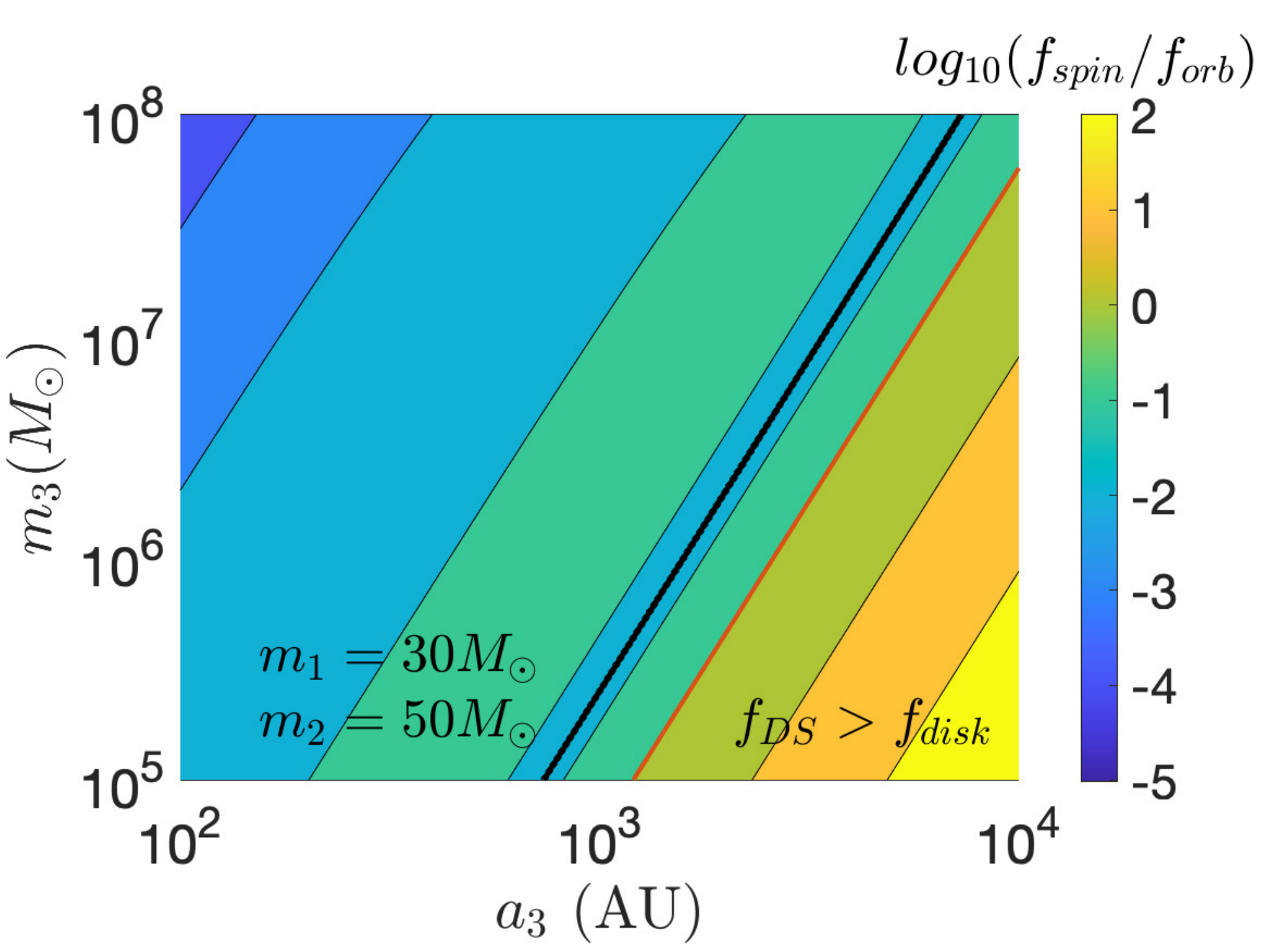}
     \includegraphics[width=0.45\textwidth]{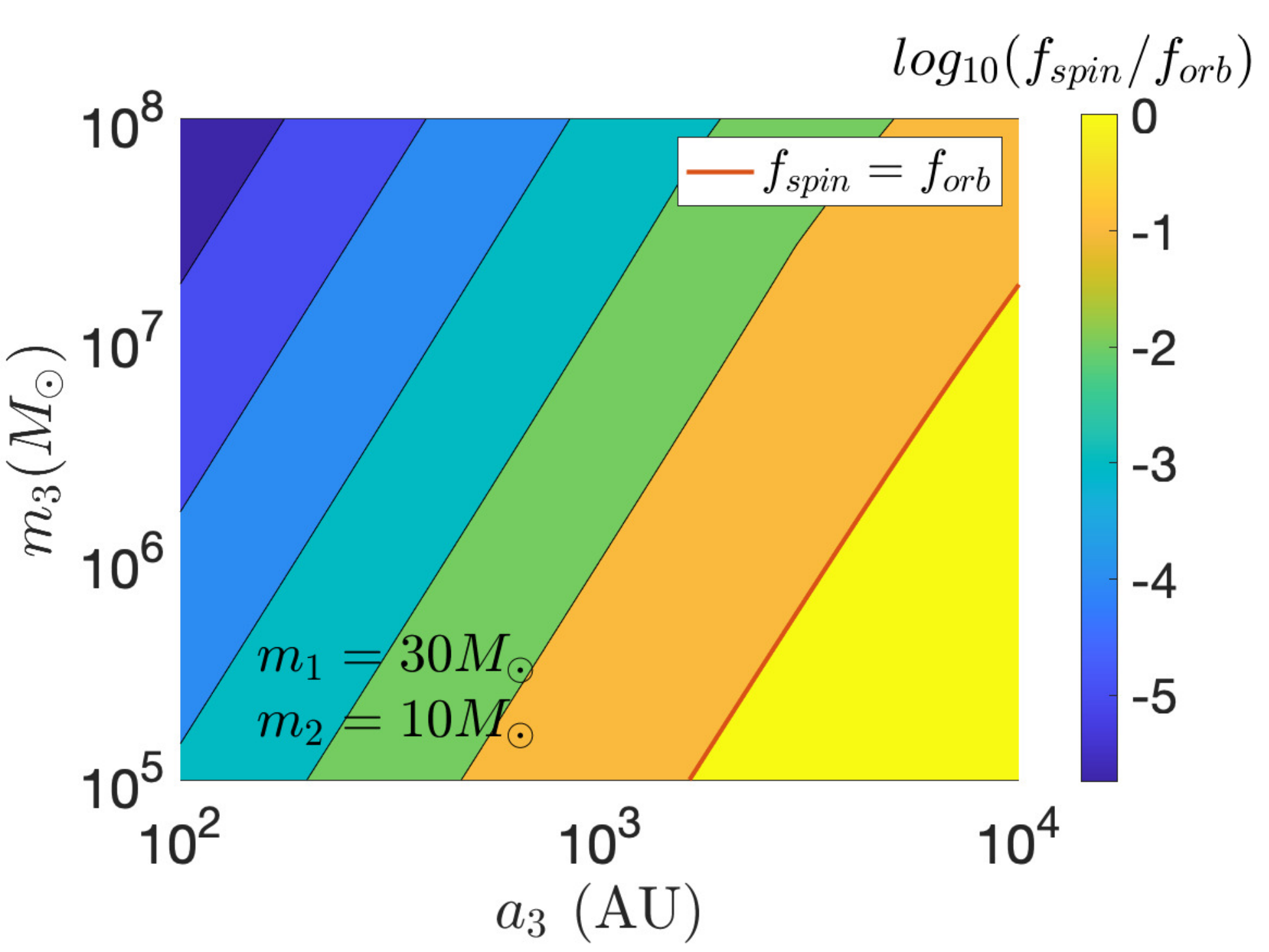}     
     \caption{Ratio of the precession frequency of the spin to that of the orbit. The \textbf{upper panel} shows the case with $m_1 = 30$\msun~sBH and $m_2 = 50$\msun~sBH, and the \textbf{lower panel} shows the case with $m_1 = 30$\msun~sBH and $m_2 = 10$\msun~sBH. The black line represents the region where the de Sitter precession frequency equals the disk induced precession frequency and the red line indicates the region where the spin precession frequency equals that of the orbit. In the lower panel, the disk precession frequency is always larger than that of the de Sitter precession frequency.} \label{fig:orbprec}
 \end{figure}

Figure \ref{fig:orbprec} shows the orbital precession rate in comparison with the spin precession rate. The x-axis represents the semi-major axis of sBH-binary orbiting around the supermassive black hole ($a_3$) while the y-axis represents the mass of the SMBH ($m_3$). The color shows the ratio of the precession rates. The upper panel includes a sBH-binary with $m_1 = 30$M${_\odot}$ and $m_2 = 50$M${_\odot}$ separated by $1$ AU, and the lower panel corresponds to $m_1 = 30$M${_\odot}$ and $m_2 = 10$M${_\odot}$ separated by $1$ AU. Similar to Figure \ref{fig:LorbLBH}, we also set $\chi = 0.7$ for illustration.

The black solid lines in the figure correspond to the region where the de Sitter spin precession frequency matches that of the disk induced spin precession frequency, assuming zero spin-orbit misalignment ($f_{DS} = f_{disk}$). The de Sitter precession is in the opposite direction to the orbital nodal regression, so the spin-orbit resonances do not occur for systems where the de Sitter precession dominates over the spin-precession (to the right of the black solid lines). The red lines in the Figure correspond to where $f_{spin} = f_{orb}$, where $f_{spin}$ represents the spin precession frequency and $f_{orb}$ represents the orbital precession frequency.

The upper panel of Figure \ref{fig:orbprec} shows that the spin precession frequency is always lower than that of the orbit when $f_{DS} < f_{disk}$. Thus, spin-orbit resonances do not occur for the higher mass sBH companion of $50$\msun. In the lower panel with a lower mass sBH companion of 10\msun, disk precession rate dominates over the de Sitter precession in all the plotted parameter space. Thus, it allows spin-orbit resonances to occur alone the red solid line. It shows that the magnitude of the orbital precession is comparable to that of the spin precession for sBH-binaries orbiting at separations of around $\sim 1000-10,000$AU for SMBH mass ranges between $10^{5-8}$\msun. sBH binaries around larger SMBH need to orbit farther from the SMBH, in order to have the orbital precession rate equal to that of the spin-precession. 

\section{Spin-orbit Resonance}
Spin-orbit resonance occurs when the spin precession rate matches that of the orbit for near coplanar orbits. For inclined orbits, the resonances occur when:
\begin{align}
    (f_{spin}/f_{orb})_{crit} &= (\sin^{2/3}{i} + \cos^{2/3}{i})^{3/2} ,
    \label{eqn:res}
\end{align}
where $i$ is the orbital inclination \citep[e.g.,][]{Ward04}. For low inclination sBH-binaries, equating the spin precession due to the circum-sBH disk (eq.~\ref{eqn:diskspin}) and the orbital nodal precession due to the SMBH (eq.~\ref{eqn:vZLK}), we determine the location of the sBH binary where the spin-orbit resonance occurs in the AGN disk:

\begin{align}
    a_{3, crit} = \Big(\frac{1}{4}\frac{a_2^6}{a_L^2}\frac{n_2}{c}\frac{m_1 m_3}{(m_1+m_2)m_
    2}\frac{\chi}{f_J}\Big)^{1/3} .
\end{align}
Substituting the expression for the Laplace radius (eq. \ref{eqn:L}), we obtain:
\begin{align}
    a_{3, crit} = \Bigg[\frac{1}{4f_J}\frac{a_2^{4}n_2}{c}\Big(\frac{a_2}{R_g}\Big)^{2/3}\frac{m_3 }{(m_1+m_2)}\Big(\frac{\chi m_1}{m_2}\Big)^{5/9}\Bigg]^{1/3} ,
    \label{eqn:a3crit}
\end{align}
where $R_{g}$ is the gravitational radius of the central sBH.


\begin{figure}[h]
\center
    \includegraphics[width=0.45\textwidth]{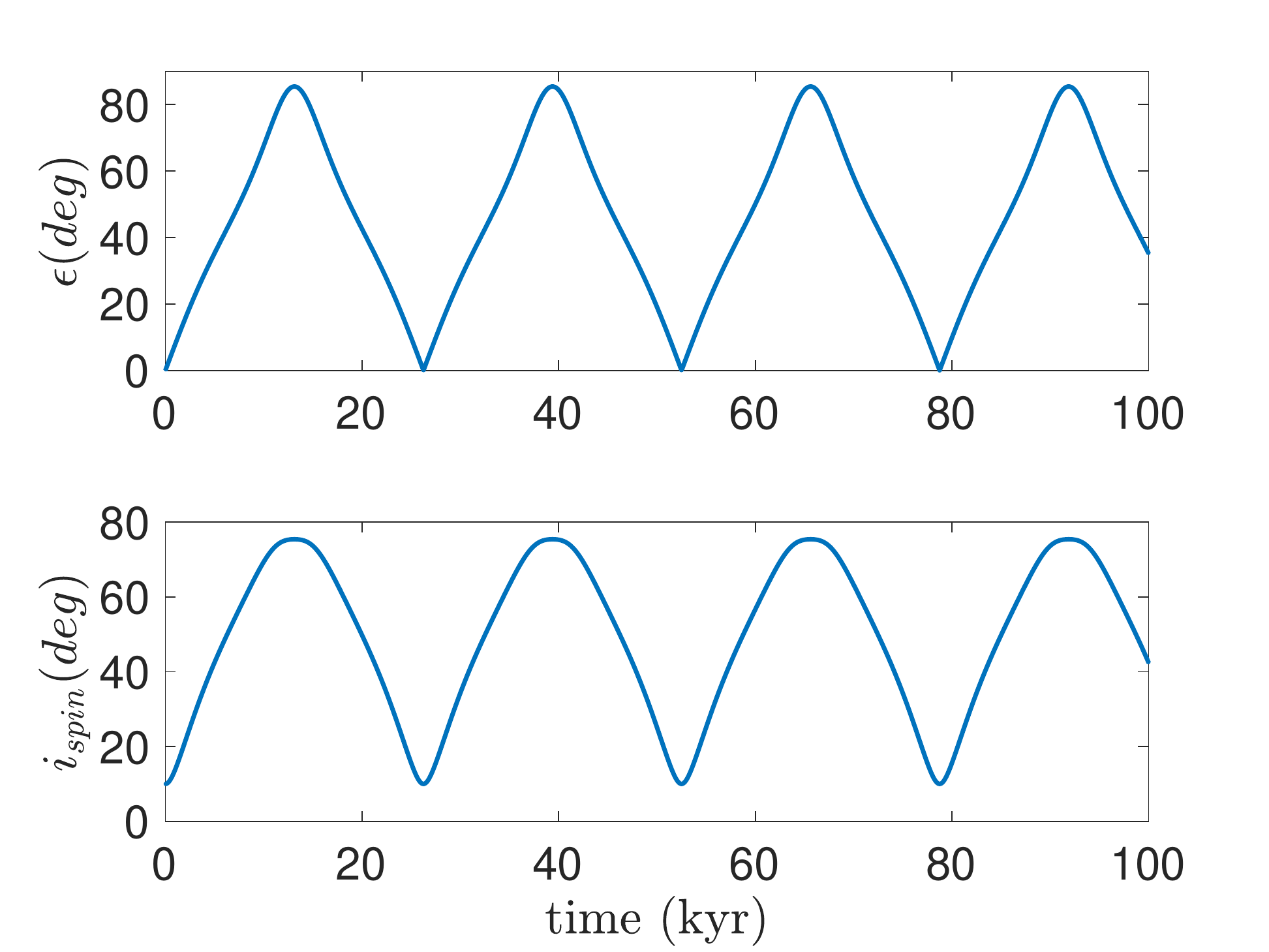}
    \caption{Upper panel: Spin-orbit misalignment (obliquity $\epsilon$) excitation of a $m_1=30$ \msun~sBH with a $m_2=10$ \msun~sBH companion, orbiting a $m_3=10^7$ \msun~SMBH.  Lower panel: spin-inclination (relative to the fixed reference plane) variations. The spin and obliquity has large variations over $\sim 10$kyr timescale due to secular spin-orbit resonances.} \label{fig:obl}
\end{figure}

To illustrate the obliquity excitation at the critical radius, we show in Figure \ref{fig:obl} the time evolution of the obliquity (upper panel) and the spin inclination relative to the orbital plane of the sBH binary around the SMBH (lower panel). We considered a $m_1 = 30$M$_{\odot}$, $\chi = 0.7$ sBH with a $10$M$_\odot$ sBH companion at a separation of $a_2 = 1$AU, orbiting around a $10^7$M$_{\odot}$ SMBH. We set the sBH binary to reside at a distance of $a_3 = 2106$AU for spin-orbit resonance, according to the expression \ref{eqn:res}. The mutual inclination of the sBH binary relative to their orbit around the SMBH is set to be $10^\circ$. Larger obliquity can be achieved when the inclination gets higher \citep{Shan18}. We integrate the spin and orbital evolution following eq.~\eqref{eqn:diskspin}, \eqref{eqn:dsspin} and \eqref{eqn:vZLK} using a fourth order Runge Kutta method. 

Figure \ref{fig:obl} shows the results of the integration. The obliquity can be excited from zero to $\sim90^\circ$ due to the spin orbit resonance, and the spin inclination relative to a fixed plane (the orbital plane of the sBH binary around SMBH $m_3$) can vary with an amplitude of $\sim70^\circ$. The spin variation timescale is $\sim 10^4$yrs, much shorter than the viscous timescale and the Bardeen-Petterson realignment timescale of the disk ($\sim 10$Myr depending on the detailed disk properties as discussed in \citet{King08, Gerosa20, Tagawa_spin20}). This allows the sBH spin not to be realigned with the outer disk immediately, even when the inner disk is not torn apart from the outer disk. Furthermore, it is shorter than the Salpeter timescale, and this allows the spin-orbit resonance to dominate over accretion effects. 

\begin{figure}[h]
\center
    \includegraphics[width=0.45\textwidth]{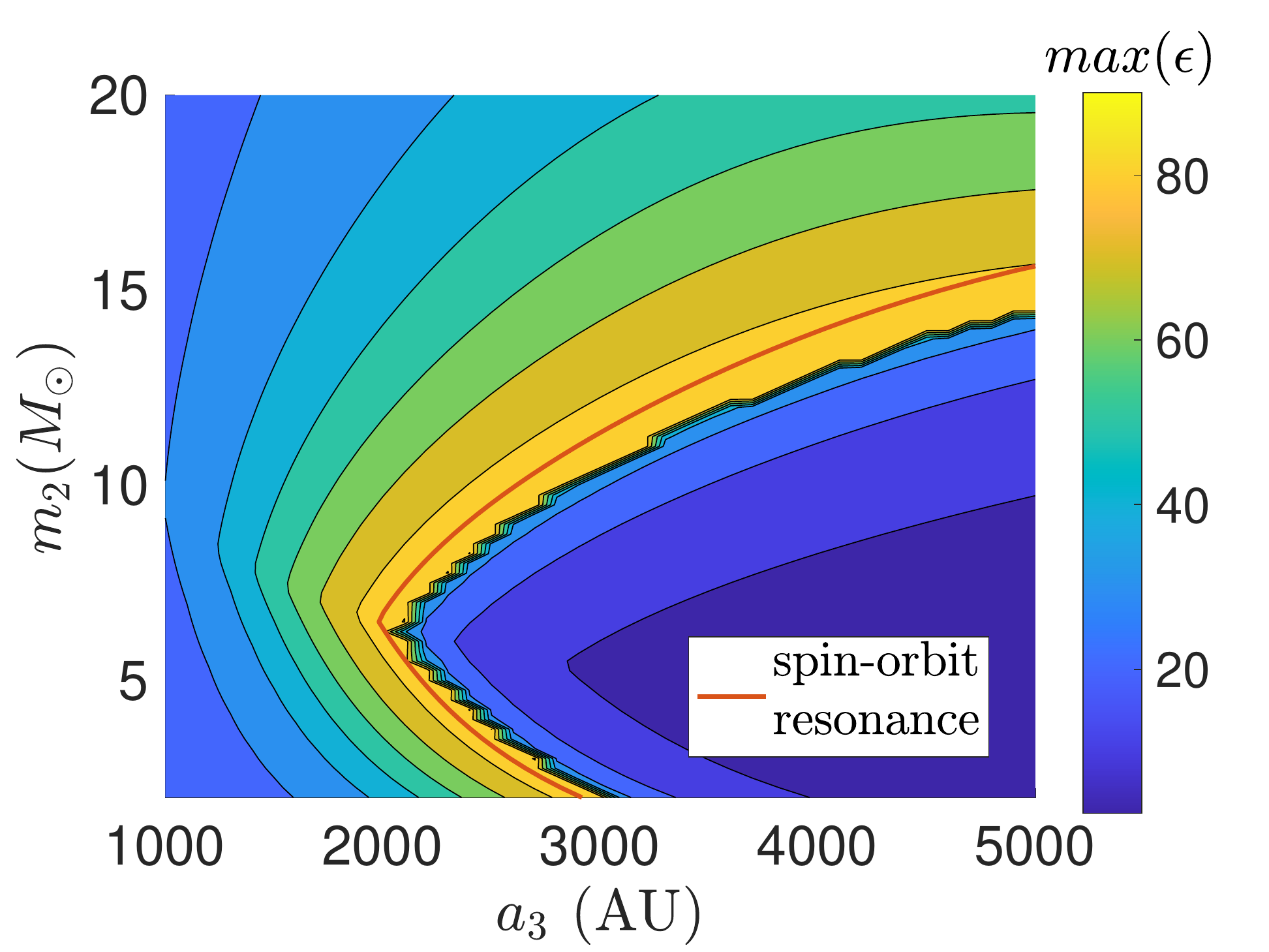}
    \caption{Maximum obliquity reached by a $30$ \msun sBH. The maximum obliquity reaches $\sim 80^\circ$ close to the red solid line, which marks the location of spin-orbit resonances based on eqn (\ref{eqn:res}). } \label{fig:maxobl}
\end{figure}

Exploring a larger parameter space, we plot in Figure \ref{fig:maxobl} the maximum obliquity reached by a $m_1 = 30$M$_{\odot}$, $\chi = 0.7$ sBH with a sBH companion of different masses at different separation from the SMBH ($m_3 = 10^7$M$_\odot$). The mutual inclination of the sBH binary relative to their orbit around the SMBH is set to be $10^\circ$. In addition, we set the separation between the sBHs to be $a_2 = 1$AU, so that most of the companions could lead to spin-orbit resonances. The x-axis of Figure \ref{fig:maxobl} shows different separation of the sBH binary from the SMBH. The red solid line in Figure \ref{fig:maxobl} shows the analytical result on the resonant location based on equation \ref{eqn:res}. It agrees well with the numerical results where the obliquity reaches maximum of $\sim 80^\circ$. When $a_3$ is small, the maximum obliquity reaches $\sim 20^\circ$, since the orbital precession is faster than the spin precession when the sBH binary is closely separated from the SMBH, and obliquity reaches twice of the orbital inclination due to the orbital precession. When the spin precession frequency becomes much faster than the orbit (large $a_3$), the spin follows the orbit and the obliquity increase is minimum.

\section{sBH-binary Migration}
So far, we assumed the sBH binary was at a fixed orbital radius around the SMBH. However, as the sBH binary interacts with the gaseous disk, it can migrate in the AGN disk and towards each other. This can lead to resonance crossing, and the spin evolution can be captured in resonance when the migration is adiabatic and the ratio of $f_{spin}/f_{orbt}$ reaches the resonant value from below following the Cassini 2 state \cite[e.g.,][]{Ward04}. On the other hand, when the ratio reaches the resonance from above, the obliquity can be kicked to large obliquity following Cassini state 1, although not captured in resonance \citep[e.g.,][]{Anderson18}.

The migration timescale in the gaseous disk is highly uncertain, depending on the disk surface density and entropy distribution \citep[e.g.,][]{chen2020}. Using the standard $\alpha$-prescription for local turbulent viscosity in AGN disks, the migration timescales for single stars/sBHs are 
estimated to be, $\tau_{\text{Type I}} \sim 10^7$ years and $\tau_{\text{Type II}} \sim 10^8$ years \citep{baruteau_binaries_2010,levin_starbursts_2007}. 
These timescales may be significantly modified by the flow geometry across the gap \citep{Kanagawa+2018},
vigorous gravito-turbulence \citep{rowther2020}, as well as the co-existence of surrounding sBHs in the AGN disk. A detailed modeling of the disk migration is beyond the scope of this study, and thus, we assume the sBH-binary and the individual sBH components migrate according to the following expressions:
\begin{align}
    a_2(t) &= a_2 (t=0) e^{-t/\tau_2}  \\
    a_3(t) &= a_3 (t=0) e^{-t/\tau_3} .
\end{align}
We assume $\tau_2\sim10^{6-8}$yr following \citet{Stone17} and $\tau_3 \sim 10^{6-8}$yr following \citep{Tagawa20} in the calculations below. 

\begin{figure}[h]
\center
    \includegraphics[width=0.45\textwidth]{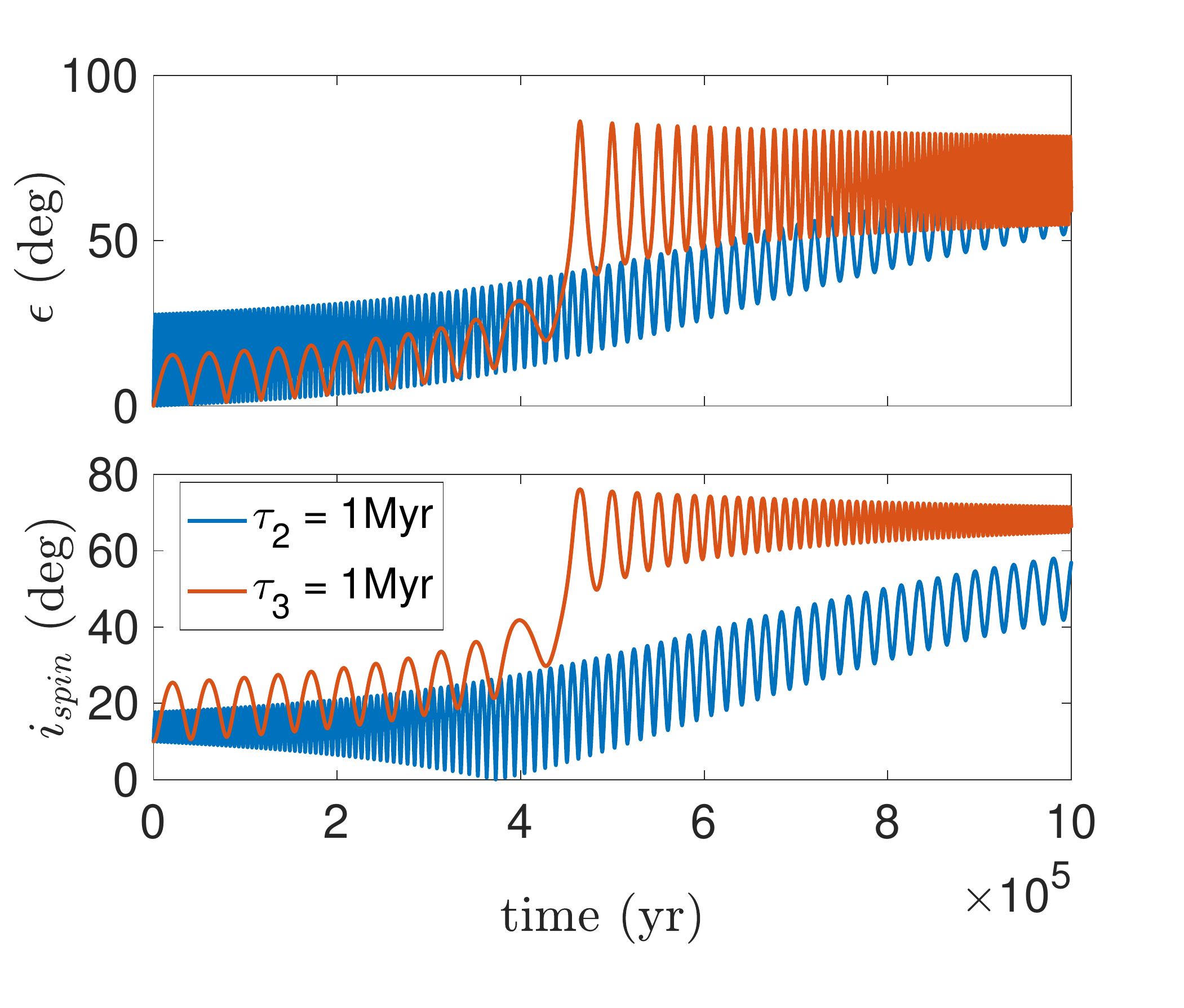}
    \caption{Obliquity and spin-inclination evolution as the sBH migrates towards each other from 2AU to $\sim 0.735$ AU (blue lines) and as the sBH-binary migrates from 4000 AU to $\sim 1471$ AU from the SMBH (red lines). The obliquity is kicked to large values faster, when $a_3$ is reduced as the sBH binary migrates towards the SMBH following Cassini state 1.} \label{fig:miga3}
\end{figure}

For illustration, Figure \ref{fig:miga3} shows the obliquity variations considering the migration of sBH binary towards the SMBH and the hardening of the sBH binary separately. We set the binary sBH of $m1=30$\msun, $m2=10$\msun orbiting around a SMBH of $m_3=10^7$\msun. The red lines correspond to the case that starts at $a_3 = 4000$AU and migrates towards  with $\tau_3 = 10^6$yr. The separation between the sBHs is $a_2=1$AU and is kept fixed in the simulation. It shows that the obliquity and the spin-inclination increases as the binary gets captured into the spin-orbit resonance at $\gtrsim 4$Myr. As the sBH migrates closer to the SMBH, the variation amplitude of the spin-axis reduces.

On the other hand, the blue lines show the case where the binary component separation starts at $a_2 = 2$AU and migrates close to each other with migration timescale of $\tau_2 = 10^6$yr. The separation from the SMBH is kept fixed at $a_3 = 2106$AU. It shows that when the binary sBH components migrate close to each other, the obliquity and the spin variation amplitude increase to large values. Note that we consider the migration of $a_2$ and $a_3$ separately in order to distinguish the effects of obliquity increase following the two different Cassini states. When the black hole binary migrate towards to the SMBH, the obliquity increase follows Cassini state 1, and the increase in obliquity is faster.

\section{Exploration of Parameter Space}
In this section, we vary all the parameters of the sBH and the SMBH to explore the sensitivity of the spin-orbit resonances on the parameters of the system configuration. Specifically, we set the spin parameters of the sBHs  to vary uniformly between $\chi = 0.1-0.9$, the masses of the sBHs ($m_1$ and $m_2$) are uniformly distributed between 2\msun to 100\msun, and the SMBH mass ($m_3$) is log-uniformly distributed between $10^{6-8}$\msun. The semi-major axis of the sBH binary is log-uniform between $a_2 = 0.1-10$AU, and the semi-major axis of the binary to the SMBH is uniformly distributed between $a_3 = 2000-5000$AU. The migration times are set to be log-uniformly distributed between $10^6$ to $10^8$yr. To investigate the amplitude in the obliquity increase of the sBHs due to the spin-orbit resonances alone, we start the sBHs all with zero obliquities and we set the inclination of the sBH-binary orbit to be 10 degrees. The larger the inclination, the higher the obliquity excitation due to the spin-orbit resonances.

Note that a population synthesis model is beyond the scope of the paper, since we only focus on the dynamics of the spin-orbit resonances. Thus, we adopt uniform/log-uniform distribution in the sampled parameters, in order to obtain an unbiased estimate on the sensitivity of the spin-orbit resonances. The results based on the uniform distribution of the parameter space can be implemented with different population synthesis models. 

\begin{figure}[h]
\center
    \includegraphics[width=0.45\textwidth]{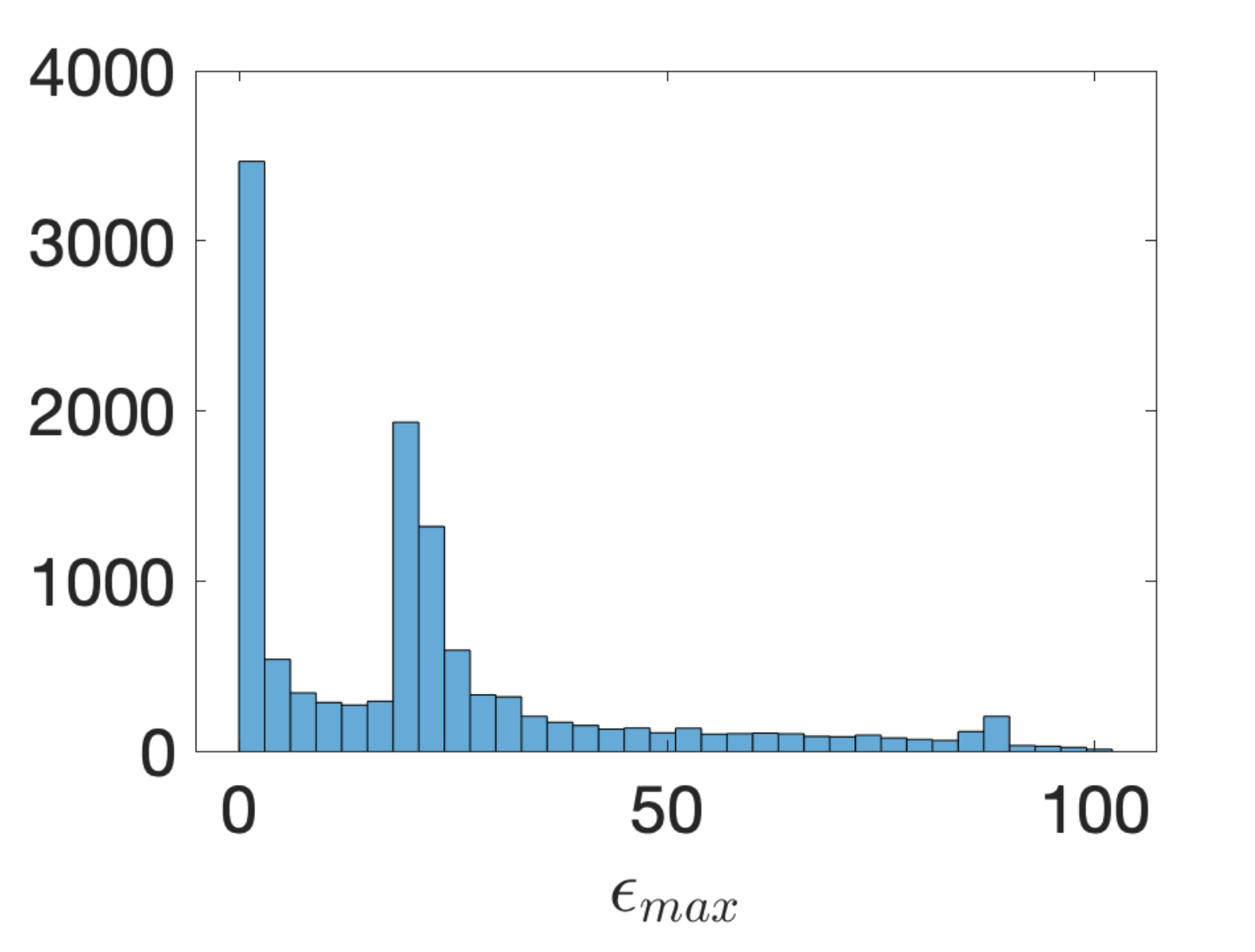}
    \caption{Histogram of the maximum obliquity reached in 10Myr. Spin-orbit resonances are common, where $\sim 50\%$ of the systems have obliquity increased above $20^\circ$ and $\sim 10\%$ of the systems reached high obliquity above $60^\circ$. The peak around $20^\circ$ is due to orbital precession.} \label{fig:hist}
\end{figure}

We simulate the evolution for a maximum of 10Myr, corresponding roughly to the realignment timescale of the spin-axis due to viscosity as discussed above, and we record the maximum obliquity reached by the sBH ($m_1$). We conducted 12,000 simulations, and removed the simulations when the sBH binary becomes unstable as the binary separation being smaller than the Hill radius of the binary relative to the SMBH ($a_3 < a_{3, Hill}$). The expression of $a_{3, Hill}$ is expressed below \citep{grishin_generalized_2017}.
\begin{equation}
a_{3,Hill}=2(3m_3/(m_1+m_2))^{1/3}a_2. \label{eqn:eqhill}
\end{equation}
Only $\sim 1\%$ of the simulations become unstable in the end of the simulation. The histogram of the maximum obliquity is shown in Figure \ref{fig:hist}. It shows that $\sim 50\%$ of the systems have obliquity increased above $20^\circ$, above that due to spin and orbital precession alone, and roughly $\sim 10\%$ of the systems have obliquity excited above $60^\circ$. This shows that spin-orbit resonances are common for the BHL disk model. A few systems reach above $90^\circ$, as a result of both spin precession and the spin-orbit resonances. The large peak at $\sim 20^\circ$ is due to orbital precession (the precession of the sBH binary around the orbit of SMBH) as the sBH spin-axis is fixed in space.

\begin{figure}[h]
\center
    \includegraphics[width=0.45\textwidth]{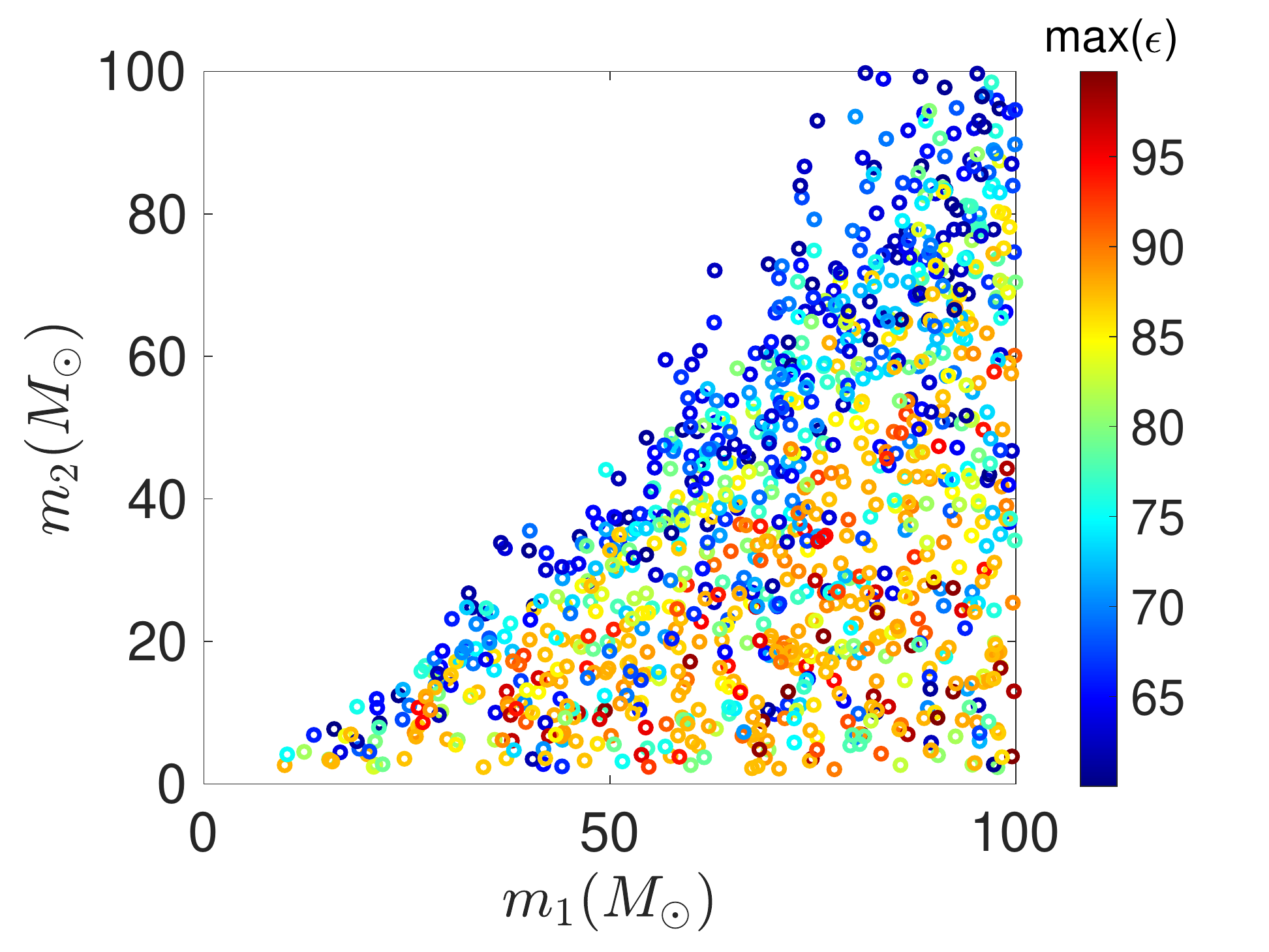}
    \includegraphics[width=0.45\textwidth]{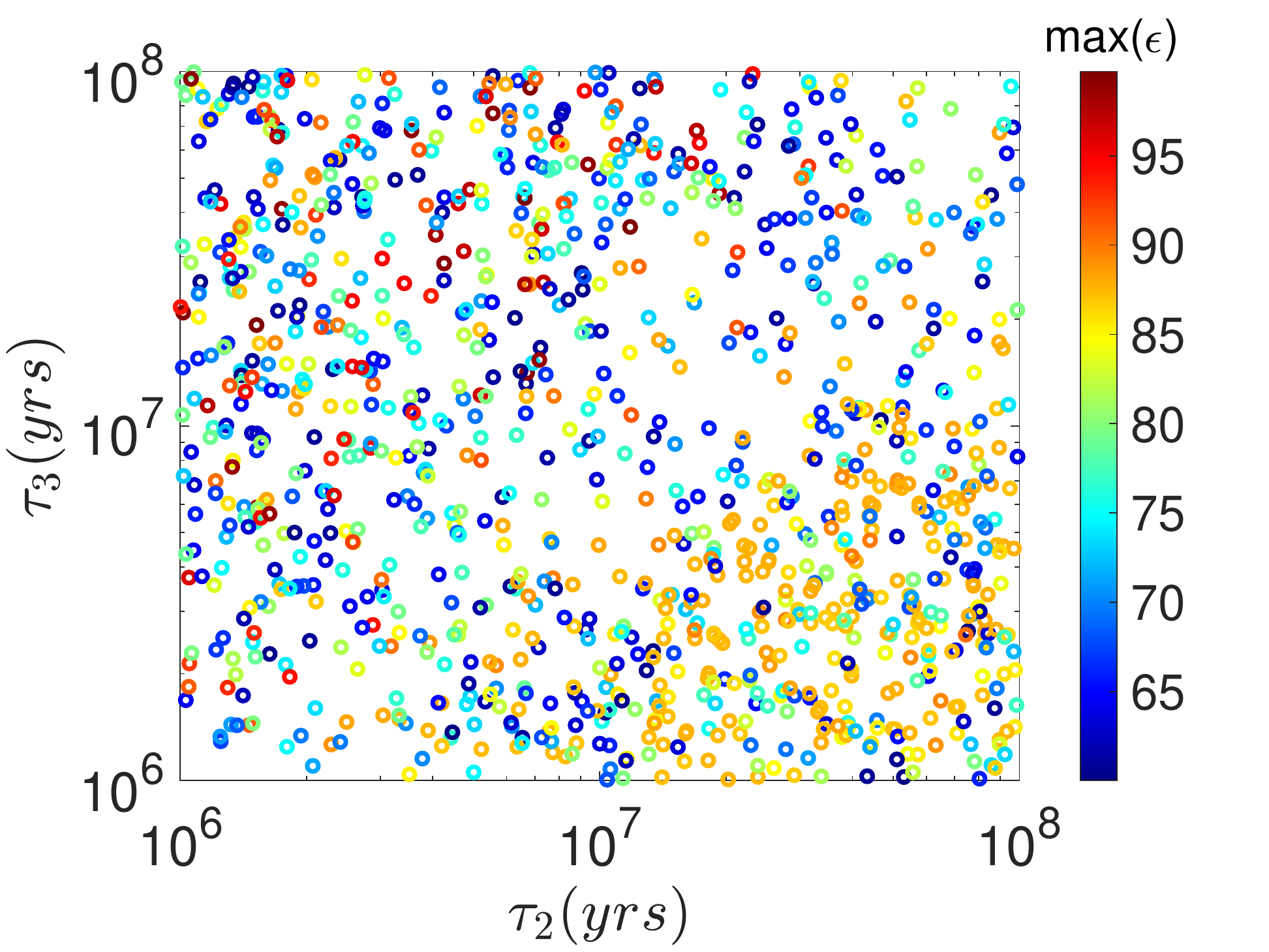}
    \caption{Scatter plot of the maximum obliquity in the plane of sBH masses, $m_1$ v.s. $m_2$ (upper panel), and in the plane of migration timescale $\tau_2$ and $\tau_3$.} \label{fig:scatter1}
\end{figure}

To illustrate the dependence of the spin-orbit resonances on the system parameters, Figure \ref{fig:scatter1} shows the scatter plot of the maximum obliquity in the plane of $m_1$ v.s. $m_2$ (upper panel), and in the plane of migration timescale $\tau_2$ and $\tau_3$ (lower panel). We only include the runs with maximum obliquity above $60^\circ$ in the plots to focus on the systems that encountered spin-orbit resonances. The upper panel shows that maximum obliquity of $m_1$ is enhanced above $60^\circ$ mostly for systems where $m_1>m_2$. This is because larger $m_1$ allows a faster disk induced $J_2$ precession to dominate over de Sitter precession, which leads to spin-orbit resonances (as discussed in section \ref{sec:spin_prec}). The sensitivity on the sBH masses leads to a wide spread in sBH binary spin-spin misalignment.

The lower panel of Figure \ref{fig:scatter1} shows that the obliquity increase is largely not sensitive to the migration rates. Looking more closely, it shows that when the sBH separation shrinks slower than the migration rate towards the SMBH ($\tau_2>\tau_3$), the maximum obliquity may only reach $\lesssim 90^\circ$, but if $\tau_2<\tau_3$, the maximum obliquity reaches $\sim 100^\circ$. This is due to the differences following the two Cassini states. Although the obliquity excitation is more abrupt following Cassini state 1 ($\tau_2>\tau_3$), the maximum obliquity extends to higher angles following Cassini state 2 as obliquity increases more smoothly within the resonance.

\begin{figure}[h]
\center
    \includegraphics[width=0.45\textwidth]{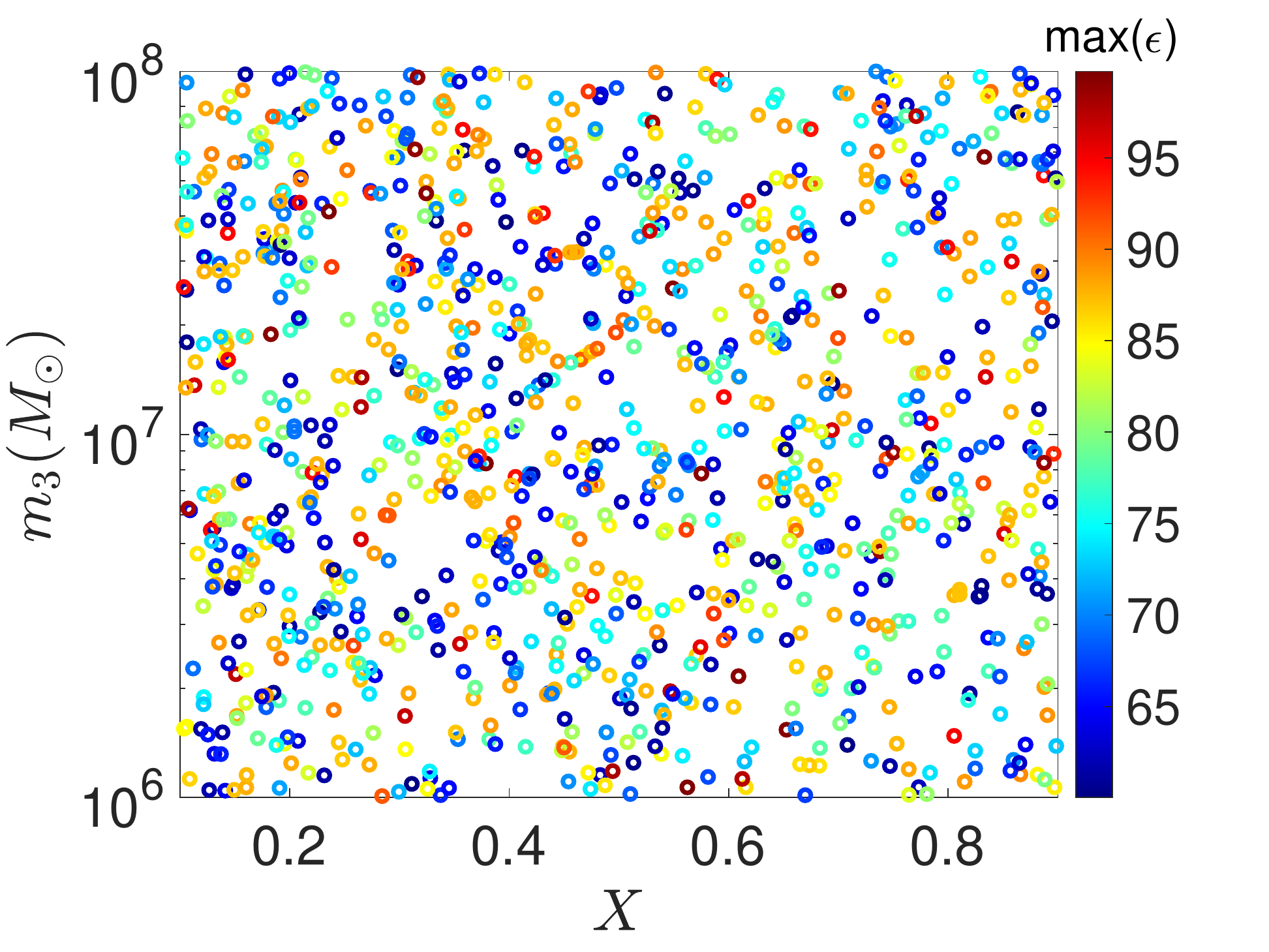}
    \caption{Scatter plot of the maximum obliquity in the plane of spin parameter ($\chi$) of $m_1$ and SMBH mass ($m_3$).} \label{fig:scatter2}
\end{figure}

In Figure \ref{fig:scatter2}, we plot the maximum obliquity in the plane of sBH ($m_1$) spin parameter ($\chi$) and SMBH mass ($m_3$). There is no clear trend in maximum obliquity as a function of the spin parameter or $m_3$. It shows that the obliquity increase is insensitive to the SBH spin parameter and the SMBH black hole mass.

\begin{figure}[h]
\center
    \includegraphics[width=0.45\textwidth]{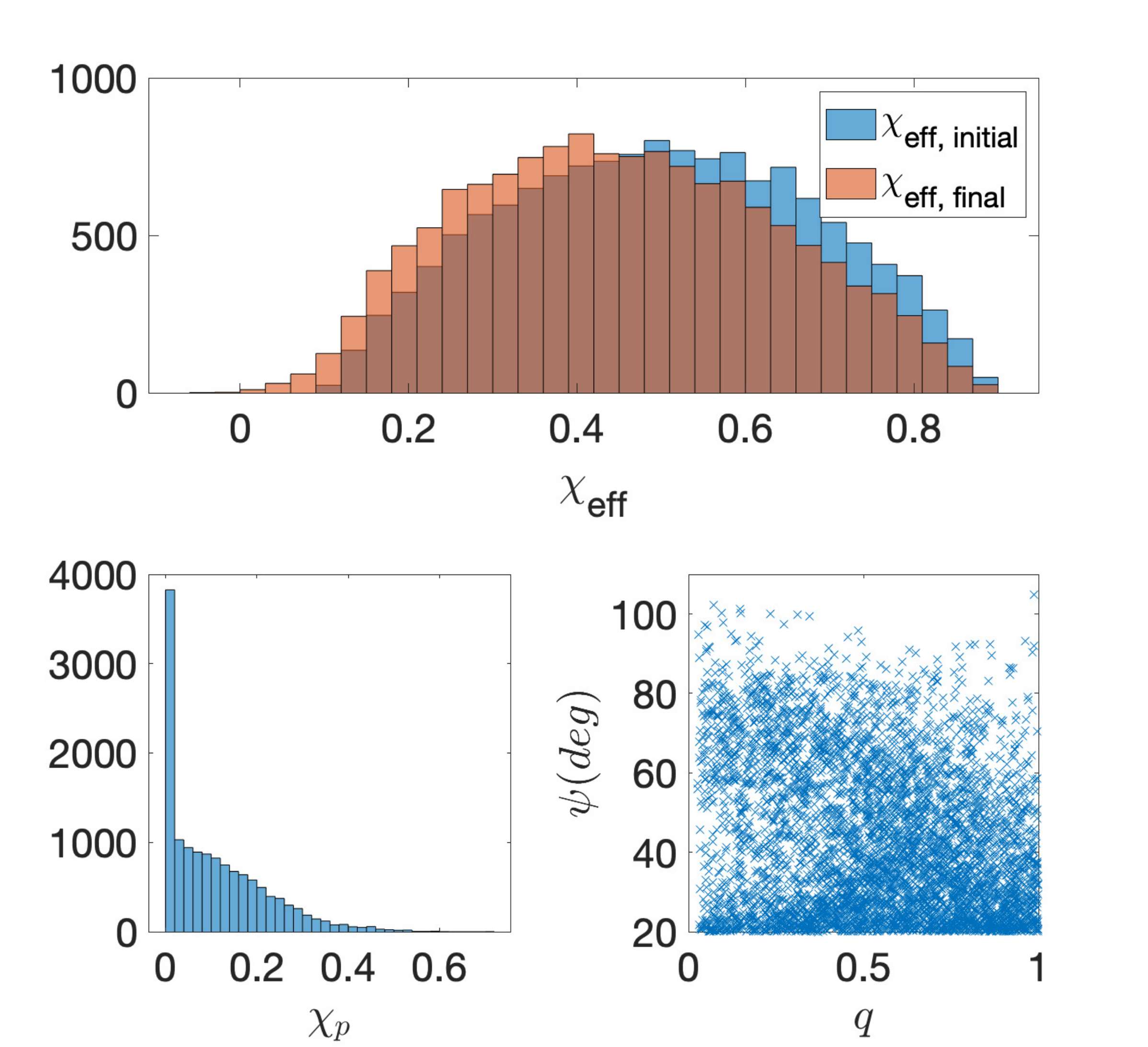}
    \caption{Distribution of the effective spin parameter, effective precession parameter and spin-spin misalignment.} \label{fig:plotchi}
\end{figure}

Evolving the spin of both $m_1$ and $m_2$, we plot in Figure \ref{fig:plotchi} the distribution of the effective spin parameter $\chi_{\rm eff}$, the effective precession parameter $\chi_p$ and the sBH spin-spin misalignment $\psi$ as a function of the mass ratio $q$ \citep{Schmidt15}. Due to sBH spin tilt, the effective spin parameter shift to smaller values, and the precession parameter increases above zero. Only a few systems obtained $\chi_{\rm eff}<0$, since spin-orbit resonances could only increase obliquity up to $90^\circ$, and obliquity can reach over $90^\circ$ due to the combined effect of spin-orbit resonances and orbital precession. $\chi_p$ can be increased up to $\sim 0.6$, and the spin-spin misalignment can reach slightly over $90^\circ$. We neglect systems with spin-spin misalignment $< 20^\circ$. spin-spin misalignment of $20^\circ$ can be due to orbital precession, as the orbits start $10^\circ$ inclined from the orbit around the SMBH.

\begin{figure}[h]
\center
    \includegraphics[width=0.45\textwidth]{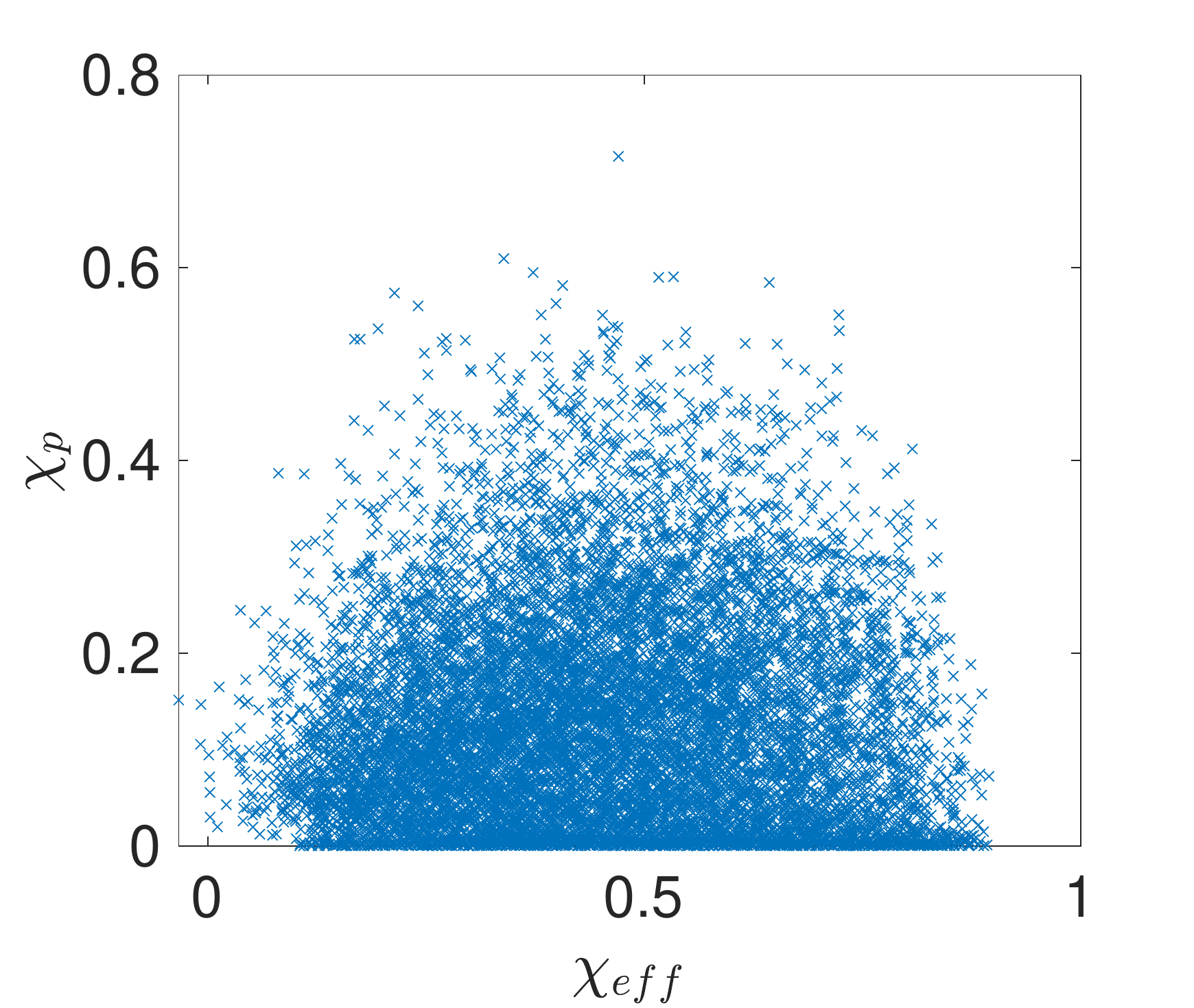}
    \caption{Scatter plot in $\chi_{\rm eff}$ and $\chi_p$.} \label{fig:chi}
\end{figure}

In the end, we show in Figure \ref{fig:chi} the scatter plot of $\chi_{\rm eff}$ and $\chi_p$. It shows that there's a wide spread in both $\chi_p$ and $\chi_{\rm eff}$, and $\chi_p$ peaks at $\sim 0.7$ when $\chi_{\rm eff}$ is around $0.5$.

\section{Conclusion and Discussions}
\label{sec:con}

In this paper, we studied changes in the spin-orientation of sBH binary components embedded in the AGN-disk due to spin-orbit resonance. Considering the torque acting on the inner disk of a sBH binary component, we find that spin-orbit resonance can drive large sBH spin obliquities (e.g., $\gtrsim 60-90^\circ$). The tilt of the sBH spin-axis due to the secular spin-orbit resonance is analogous to the tilt of Saturn's axis considering the effects of the satellites and the close-in circumplanetary disk \citep{Ward04, Rogoszinski20}. However, we note that this only occurs for sBHs surrounded by very massive disks following super Eddington BHL accretion, and an Eddington-limited disk may cause the quadrupole moment to be too low for resonances to occur.


The spin of the sBHs precess due to two different mechanisms: de Sitter precession due to the spacetime curved by the sBH companion, and the disk induced $J_2$ Newtonian gravitational potential. The direction of the precession are opposite to each other, and the disk $J_2$ precession is in the same direction as the orbit and can thus lead to spin-orbit resonances. However, secular spin-orbit resonances do not occur when de Sitter precession dominates. We find that spin-orbit resonance is more likely to occur for more massive sBHs with smaller sBH companions, so that the de Sitter precession rate is relatively slow. In addition, a massive circum-sBH disk is required to have large enough $J_2$-precession that allows for the presence of spin-orbit resonances. Assuming an $\alpha$-disk model with Bondi-Hoyle-Lyttleton accretion, the resonances take place for sBH binary semi-major axis of $a_2 \sim 1$AU, and for large sBH mass with a lower mass sBH companion. The de Sitter spin-precession rate dominates over the disk-induced spin precession for harder binaries, which cannot lead to spin-orbit resonances. 

Orbiting around a SMBH of $10^7$ \msun, the sBH binary needs to reside at around $\sim 1000$AU away from the SMBH so that the nodal precession rate due to the SMBH matches that of the spin precession and lead to spin-orbit resonances. The migration of the sBHs can lead to resonance capture. When the migration of the sBH-binary towards the SMBH is faster than the shrinking of the sBH orbit, obliquities can be kicked to large values following Cassini State 1. The spin-orbit resonances can excite obliquity to $\sim 90^\circ$ and lead to a wide spread of spin effective parameters and sBH binary spin-spin misalignment for unequal mass sBHs. 

Using 12000 MC simulations, with uniform distribution in sBH masses, spin parameters and separation from the SMBH ($a_3$), and log-uniform ditribution in SMBH masses and migration rates, we find that spin-orbit resonances occur in $\sim 50\%$ of the simulations, with $\sim 10\%$ of them having spin-orbit misalignment excited over $60^\circ$. The spin-orbit resonances are more likely to occur when the sBH companion mass is lower than that of the primary.

As a caveat, we point out that the secular spin-orbit resonances only occur for massive circum-sBH disk or with very low mass companion, so that the circum-sBH disk provides a large enough $J_2$ potential to compete with the de Sitter spin precession. This can take place for super-Eddington accretion circum-sBH disks (e.g., with Bondi-Hoyle-Lyttleton accretion), which may occur for non-spherically symmetric accretion, and can be quenched by radiation pressure episodically. However, the fast spin-variations (with kyr timescales) may still allow the sBH obliquities to be excited during the super Eddington accretion cycle. Beyond sBH-binaries, such spin flips can also occur for IMBH-sBH binaries, as well as sBH-star binaries, in less massive disks. In fact, sBH-star binaries can have a large occurrence rate since they are more easily to be captured due to tidal interactions. Moreover, forced accretion of sBH-binaries inside the envelope of massive stars can also allow massive disks to exist longer, and possibly enhance the rate of spin-orbit resonances. 

We note that we restricted our focus on the effects due to the spin-orbit resonance in this paper as a proof-of-concept, and  illustrated the effects using a few examples to investigate the qualitative behavior. Exploring the implications on LIGO/VIRGO data with population synthesis in a large parameter space is beyond the scope of the paper, and will be addressed in a follow-up study. Beyond secular spin-orbit resonances, subsequent evolution of the sBH spin-orbit misalignment due to interactions with stellar flyby/sBHs can greatly enhance the inclination of the orbits with respect to the disk and increase the spin-orbit misalignment of the sBHs \citep{Tagawa_spin20}. In addition, gas accretion onto the sBHs in a turbulent disk can also change the spin-orientation of the sBHs. These effects can allow $\chi_{\rm eff}$ to be centered around zero and agree with observational results, as discussed in \citet{Tagawa_spin20,Tagawa21,Secunda20,McKernan20,McKernan22}. In the future, astrophysical GW echos can identify sources situated in the vicinity of a SMBH and further constrain whether the obliquity increases could be due to spin-orbit resonance \citep{Gondan21}.

\begin{acknowledgments}
The authors thank the referee for comments that significantly improved the quality of the paper, and the authors thank Yubo Su and Zoltan Haiman for helpful discussions. GL and HB are grateful for the support by NASA 80NSSC20K0641 and 80NSSC20K0522. This work used the Hive cluster, which is supported by the National Science Foundation under grant number 1828187.  This research was supported in part through research cyberinfrastrucutre resources and services provided by the Partnership for an Advanced Computing Environment (PACE) at the Georgia Institute of Technology, Atlanta, Georgia, USA.
\end{acknowledgments}

\appendix

\bibliography{sample631.bib}
\bibliographystyle{aasjournal}

\end{document}